\documentclass[letterpaper,aps,prb,twocolumn,floatfix,amsfonts,amssymb,superscriptaddress]{revtex4-1}

\usepackage{amsmath}
\usepackage{hyperref}
\usepackage{graphicx}
\usepackage[dvipsnames]{xcolor}

\hypersetup{
  colorlinks   = true, %Colours links instead of ugly boxes
  urlcolor     = Blue, %Colour for external hyperlinks
  linkcolor    = Blue, %Colour of internal links
  citecolor   = Blue %Colour of citations
}

\DeclareMathOperator{\diag}{diag}
\DeclareMathOperator{\sgn}{sgn}
\renewcommand{\Re}{\operatorname{Re}}
\renewcommand{\Im}{\operatorname{Im}}
\pdfoutput=1

\begin{document}
\title{Signatures of time-reversal-invariant topological superconductivity in the Josephson effect}
\author{Ehren Mellars}
\affiliation{School of Physics and Astronomy, University of Birmingham, \\ Edgbaston, Birmingham, B15 2TT, United Kingdom.}
\author{Benjamin B\'eri}
\affiliation{School of Physics and Astronomy, University of Birmingham, \\ Edgbaston, Birmingham, B15 2TT, United Kingdom.}
\date{July 2016}

\begin{abstract}
	\indent For Josephson junctions based on $s$-wave superconductors, time-reversal symmetry is known to allow for powerful relations between the normal-state junction properties, the excitation spectrum, and the Josephson current.  Here we provide analogous relations for  Josephson junctions involving one-dimensional time-reversal-invariant topological superconductors supporting Majorana--Kramers pairs, considering both topological--topological and $s$-wave--topological junctions. Working in the regime where the junction is much shorter than the superconducting coherence length, we obtain a number of analytical and numerical results that hold for arbitrary normal-state conductance and the most general forms of spin-orbit coupling. The signatures of topological superconductivity we find include the fractional ac Josephson effect, which arises in topological--topological junctions provided that the energy relaxation is sufficiently slow.  We also show, for both junction types, that robust signatures of topological superconductivity arise in the dc Josephson effect in the form of switches in the Josephson current due to zero-energy crossings of Andreev levels. The junction spin-orbit coupling enters the Josephson current only in the topological--topological case and in a manner determined by the switch locations, thereby allowing quantitative predictions for experiments with the normal-state conductance, the induced gaps, and the switch locations as inputs.
\end{abstract}

\maketitle
\section{Introduction} \label{sec:intro}

Josephson junctions involving Majorana fermions\cite{KitaevPU2001, FuKanePRL2008, AliceaRPP2012, BeenakkerARCP2013} are under intensive theoretical\cite{FuKanePRB2009, IoselevichPRL2011, HeckPRB2011, JiangPRL2011, SJosePRL2012, DominguezPRB2012, ZhangKaneMele2013, ChungPRB2013, KTLawPRX2014, PikulinPRB2016, IoselevichPRB2016,  MarraPRB16, LiuarXiv2016,NavaarXiv2016} and experimental\cite{Rokhinson2012fractional, DengNL2012, Pribiag2015edge, WiedenmannNatComms2016, BocquillonNatureNano2016} investigation for the promising routes they provide towards demonstrating topological superconductivity\cite{SchnyderPRB2008, QiPRL2009, AliceaRPP2012, BeenakkerARCP2013} and as potential building blocks towards topological quantum computation.\cite{FuKanePRL2008,BeenakkerARCP2013,HasslerNJoP2010, HyartPRB2013} In the most frequently studied class of systems, Majorana fermions are nondegenerate zero-energy end states in hybrid devices realising one-dimensional (1D) effectively spinless $p$-wave superconductors.\cite{KitaevPU2001, FuKanePRB2009, LutchynPRL2010, OregPRL2010, Duckheim2011, NadjPerge2013, NadjPerge2014} A number of concrete platforms exist to realise this time-reversal symmetry breaking (so-called class D\cite{AltlandZirnbauerPRB1997}) topological superconductivity, all of which use the proximity effect to combine $s$-wave superconductors, strong spin-orbit coupling (e.g., via nanowires,\cite{LutchynPRL2010, OregPRL2010} topological insulators,\cite{FuKanePRB2009} or the superconductor itself\cite{Duckheim2011, NadjPerge2013, NadjPerge2014}) and magnetic fields (e.g., via Zeeman coupling\cite{LutchynPRL2010, OregPRL2010, AliceaPRB2010} or ferromagnetism\cite{FuKanePRB2009, Duckheim2011, NadjPerge2013, NadjPerge2014}).

The past years have seen a rapidly increasing interest in realising\cite{NakosaiPRL2012, NakosaiPRL2013, KTLawPRB2012, ZhangKaneMele2013, SunPRB2014, HaimPRB2014, KlinovajaPRB14a, KlinovajaPRB14b, SchradePRL15, HaimPRB2016, EbisuPTEP2016} and detecting\cite{ZhangKaneMele2013, ChungPRB2013, KTLawPRX2014, LiPRL2016, PikulinPRB2016, KimPRB2016,LiuarXiv2016} time-reversal-invariant (so-called class DIII\cite{AltlandZirnbauerPRB1997}) analogues of such Majorana fermion systems.  In 1D, class DIII topological superconductors host a Kramers pair of Majorana fermions at each end that can be combined into a zero-energy fermion end mode with  anomalous  time-reversal properties.\cite{QiPRL2009} The proposed platforms for realisation again include hybrid devices based on spin-orbit coupling\cite{KTLawPRB2012, ZhangKaneMele2013, HaimPRB2014, HaimPRB2016, LiuarXiv2016} and the superconducting proximity effect,\cite{NakosaiPRL2012, NakosaiPRL2013, ZhangKaneMele2013, SunPRB2014, LiuarXiv2016} albeit now with unconventional (but nontopological) superconductors (e.g., iron-based superconductors with $s_\pm$-wave pairing\cite{ishida2009extent,mazin2010superconductivity,UmezawaPRL2012}). These hybrids realise spinful effectively $p$-wave pairing with the admixture of a smaller $s$-wave pairing component.  

\begin{figure}[b]
\includegraphics[width=0.85\columnwidth]{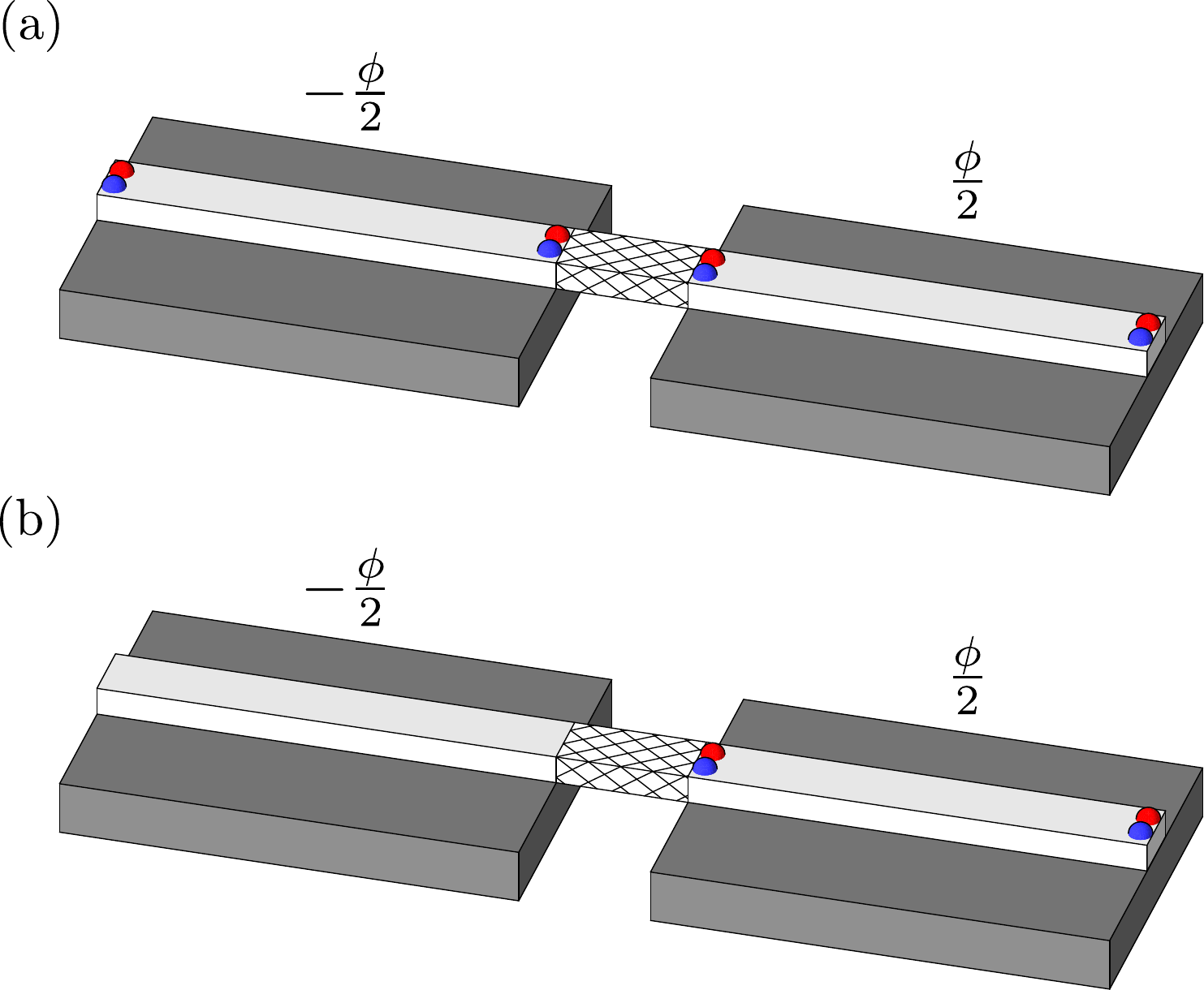}
\caption{(Colour online) Majorana--Kramers setups for 1D Josephson junctions with  a phase difference $\phi$ across the junction. The superconductors (dark grey slabs) each have a 1D, spin-orbit coupled quantum wire (light grey blocks) deposited on top of them. By the proximity effect, superconductivity is induced in each quantum wire. The wires with Majorana--Kramers pairs (red and blue hemispheres) display induced topological superconductivity, arising from coupling to unconventional (e.g., $s_\pm$-wave) superconductors. The normal region (e.g., tunnel barrier, point contact, disordered metal) between the superconductors is indicated by cross-hatching. 
}
\label{fig:setups}
\end{figure}

In this paper, we study Josephson junctions involving 1D class DIII topological superconductors. We obtain the energy spectrum and the consequent Josephson current in terms of the normal-state scattering properties, requiring only that the junction be much shorter than the superconducting coherence length (i.e. the short-junction limit). Our results apply for a number of physically relevant situations including disordered junctions, or junctions of intermediate transparency that are away from both the tunnelling and the highly transparent limits considered in previous works.\cite{ZhangKaneMele2013, ChungPRB2013, KTLawPRX2014}

We obtain results for two types of setups (Fig.~\ref{fig:setups}): junctions between two topological superconductors [Fig.~\ref{fig:setups}(a)] and junctions between an $s$-wave superconductor and a topological superconductor [Fig.~\ref{fig:setups}(b)]. 
For the first type of setup [Fig.~\ref{fig:setups}(a)], we study the conditions under which a time-reversal-invariant generalisation\cite{ZhangKaneMele2013,KTLawPRX2014} of the so-called fractional Josephson effect\cite{KitaevPU2001, KwonEPJB2004, FuKanePRB2009, LutchynPRL2010, OregPRL2010, IoselevichPRL2011, KTLawPRB2011} can arise. This hallmark signature in class D topological superconductors refers to the appearance of a $4\pi$ periodic current-phase relation, replacing the more conventional $2\pi$ periodic one as Majorana fermions enable charge transfer in terms of electrons instead of Cooper pairs.
In addition to establishing the form in which such behaviour can appear in class DIII topological Josephson junctions with generic junction properties, we also consider the role of the characteristic timescale for energy relaxation. The latter aspect, to the best of our knowledge, has so far not been considered; however, as we show, it presents an important channel for the loss of $4\pi$ periodicity.

For the second type of setup [Fig.~\ref{fig:setups}(b)], we investigate how the anomalous time-reversal properties of the fermion $f$, built from the two interface Majoranas, appear in the features of the Josephson current. Here general considerations\cite{ChungPRB2013} show that there is an anomalous, time-reversal protected contribution that gives nonzero current at $\phi=n\pi$ (where $n$ is an integer) with the sign set by the parity of $f$. To positively establish the origin of this contribution in an experimental system (e.g., that the current at $\phi=n\pi$ is not due to broken time-reversal invariance\cite{KwonEPJB2004}), however, a theory for the current-phase relationship is needed that accounts for other, nonanomalous contributions and which holds for generic junctions. Such a theory may also be useful in scenarios in which access to changes in $\phi$ are more readily available than to $\phi$ itself; for example, the ac Josephson effect where the phase sweep speed is controlled by the voltage across the junction (i.e. $\dot{\phi}=2eV/\hbar$). In this paper we provide such a theory.

The signatures mentioned above are for junctions with a conserved fermion parity. We also investigate, for both setups, the case in which the fermion parity is allowed to relax (i.e. the dc Josephson effect regime). For all the regimes to be considered, we compare topological junctions to their nontopological counterparts (i.e. without Majorana fermions) in order to search for unique signatures of topological superconductivity.

The rest of this paper is organised as follows.  We start, in Sec.~\ref{sec:Hamiltonian}, by describing 1D class DIII topological superconductors in terms of the Bogoliubov--de Gennes Hamiltonian  near the  Fermi points. This will allow us, in Sec.~\ref{sec:Smatrices}, to obtain the Andreev reflection matrices of normal--topological superconductor interfaces, and thus to formulate a scattering matrix description for our systems. In Sec.~\ref{sec:AndreevBS} we investigate the bound-state spectrum supported by topological--topological (Sec.~\ref{sec:ps-ps}) and $s$-wave--topological (Sec.~\ref{sec:s-ps}) Josephson junctions. Then in Sec.~\ref{sec:JosephsonCurrent} we calculate the Josephson currents for our junction setups and contrast the results with the corresponding nontopological cases. To test our scattering matrix predictions, in Sec~\ref{sec:numerics}, we compare our results to numerical simulations of a lattice model of time-reversal-invariant Josephson junctions. Finally, in Sec.~\ref{sec:concl} we present our conclusions.

\section{Hamiltonian description of topological superconductors}
\label{sec:Hamiltonian}

In mean field theory, the excitations of superconductor systems can be described in an effectively single-particle picture in terms of the so-called Bogoliubov--de Gennes Hamiltonian,\cite{DeGennesBook} which is a $2\times2$ matrix in electron-hole space
\begin{equation}
	H_{\text{BdG}}=\left(\begin{array}{cc}
	h_\mathrm{e} & \Delta\\
	\Delta^{\dagger} & h_\mathrm{h}
	\end{array}\right).
	\label{eq:HBdG}
\end{equation}
Here $h_\mathrm{e}$ and $h_\mathrm{h}$ are the single-particle Hamiltonians for electrons and holes, respectively, and $\Delta$ is the superconducting pair potential.
Each of the blocks of Eq.~\eqref{eq:HBdG} acts on spin-$1/2$ electrons and we have $h_\mathrm{h}=-\mathcal{T}h_\mathrm{e}\mathcal{T}^{-1}$ where $\mathcal{T}=i\sigma_2K$ is the operator for time reversal with Pauli matrices $\sigma_j$ acting in spin-space and complex conjugation operator $K$. This choice of $h_\mathrm{h}$ corresponds to a basis that makes spin-rotation properties manifest: in a spin-rotation symmetric system, all blocks of $H_\text{BdG}$ are proportional to the identity matrix in spin-space.

A Josephson junction is composed of two superconductors linked together by a normal region with a phase difference $\phi$ across the junction. We will describe such junctions, with normal regions $|x|\leq\frac{l}{2}$, using the step-function model\cite{LikharevRMP1979}
\begin{equation}
	\Delta=\begin{cases}
		\Delta_{\mathrm{L}} e^{-i\frac{\phi}{2}} & x<-\frac{l}{2},\\
		0 & |x|\leq \frac{l}{2},\\
		\Delta_{\mathrm{R}} e^{i\frac{\phi}{2}} & x>\frac{l}{2}.
\end{cases}
\end{equation}

To describe topological superconductors in our setups, we focus on the case in which $h_\mathrm{e}$ describes a spinful system that is, at least in the superconductors and at the normal--superconductor interfaces, effectively 1D (e.g., the spin-orbit coupled nanowire in the hybrid proposals\cite{NakosaiPRL2013, KTLawPRX2014}) with a single conducting channel. The two spin components give rise to two right-moving and two left-moving Fermi points. Class DIII topological superconductivity arises\cite{ZhangKaneMele2013}  when the spectrum acquires superconducting gaps $\Delta_\pm$ of opposite signs at the two right-moving Fermi points. The gaps at the left-moving pair of Fermi points follow by time-reversal symmetry.  For our purposes, it is sufficient to focus on the physics in the vicinity of the Fermi points. In this case, in addition to spin, $h_\mathrm{e}$ acquires a further $2\times 2$ block structure for left- and right-movers. We have 
\begin{subequations}
	\begin{align}
		h_{\mathrm{e}/\mathrm{h}}&=\pm i\hbar\, \diag\left(v_{+},v_{-},-v_{-},-v_{+}\right)\partial_{x}\label{eq:ehlin},\\
		\Delta_{o}&=\diag\left(\Delta_{+o},\Delta_{-o},\Delta_{-o},\Delta_{+o}\right)\label{eq:pairinglin},
	\end{align}
	\label{eq:HBdGlin}% ending in % needed to prevent indentation afterwards http://tex.stackexchange.com/questions/54032/text-following-subequations-is-slightly-indented-if-a-label-is-used
\end{subequations}
where the entries are ordered from the rightmost to the leftmost Fermi point. In Eq.~\eqref{eq:ehlin}, $v_+$ and $v_-$ are the Fermi velocities where the upper sign refers to the electron Hamiltonian and the lower to the hole Hamiltonian. In Eq.~\eqref{eq:pairinglin}, $o=\mathrm{L},\mathrm{R}$ refers to the left/right superconductor and, without loss of generality, we will take $\Delta_{+o}>0$. Moreover, by parameterising the pairings as $\Delta_\pm=\Delta_s \pm \Delta_p$, the system can be viewed as having a time-reversal-invariant $p$-wave pairing $\Delta_p$ with the admixture of a smaller $s$-wave pairing $\Delta_s$. In line with a number of concrete proposals describing systems conserving the $z$-component of spin, one can view $v_\pm$  as the right-/left-moving spin-$\uparrow$ velocities and $\Delta_\pm$ as the pairing at the corresponding Fermi points. (The complementary spin-$\downarrow$ quantities follow via time reversal.) The Hamiltonian~\eqref{eq:HBdGlin} is, however, more general and holds in the absence of a conserved spin component; it can be used to describe the superconductors in both setups we consider. For the topological--topological junction, we consider the gap-symmetric case $\Delta_\mathrm{L}=\Delta_\mathrm{R}$, while for the $s$-wave--topological junction we impose no such requirement.

\section{Scattering matrix description of (topological) Josephson junctions}
\label{sec:Smatrices}

For nontopological, singlet superconductor-based, Josephson junctions in the short-junction limit, time-reversal invariance is known to allow for elegant relations to the normal-state scattering properties,\cite{HaberkornPSS1978, BeenakkerPRL1991,TanakaPRB96,TanakaPRB97,Riedel98,KashiwayaRPP00} such as Beenakker's formula for $s$-wave junctions\cite{BeenakkerPRL1991}
\begin{align}
	E=\sqrt{1-\tau\sin^2\left(\phi/2\right)},
	\label{eq:BeenakkerFormula}
\end{align}
expressing the Andreev (i.e. subgap) bound-state energies in terms of the normal-state transmission probability $\tau$ at the Fermi energy. As we show below, a number of analogous exact relations exist in the topological case. The starting point for establishing these are the Andreev reflection matrices at  normal--(topological) superconductor interfaces.

\subsection{Andreev reflection matrices for topological superconductors}

Excitations with energies below the superconducting gap can be described in terms of Andreev reflections at the superconductor--normal (SN) interfaces. These Andreev reflections are described by the Andreev reflection matrices $r_\mathrm{eh}$ and $r_\mathrm{he}$, which contain amplitudes for a hole reflecting into an electron and vice versa, respectively.

We now use Eq.~\eqref{eq:HBdGlin} as a starting point to derive the Andreev reflection matrices at SN interfaces, which will serve as key ingredients for formulating the scattering matrix description of our systems. The superconductors are 1D, forming single transport channels that carry electrons and holes of a particular spin, leading to four subchannels overall. The key assumption that we make is that the four electron-hole Dirac cones at the four Fermi points remain uncoupled for clean SN interfaces. This so-called Andreev approximation\cite{AndreevZhETF1964} holds when the superconducting coherence length is much longer than the inverse of the separation between neighbouring Fermi points or, when the bulk superconducting system conserves a spin component, between the left- and right-moving Fermi points.

The approach to deriving the Andreev reflection matrices is essentially the same as for the Josephson junction composed of $s$-wave superconductors, as treated by Beenakker.\cite{BeenakkerPRL1991} The Andreev reflection matrices take the form
\begin{align}
	r_\mathrm{eh}=
	\begin{bmatrix}
		\kappa_{1\mathrm{L}} & 0 \\
		0 & \kappa_{2\mathrm{R}}
	\end{bmatrix}, \quad
	r_\mathrm{he}=
	\begin{bmatrix}
		\kappa_{2\mathrm{L}} & 0 \\
		0 & \kappa_{1\mathrm{R}}
	\end{bmatrix},
	\label{eq:andreev0}
\end{align}
where we have introduced the subblocks
\begin{align}
	\kappa_{1o}=
	\begin{bmatrix}
		\alpha_{-o} & 0 \\
		0 & \alpha_{+o}
	\end{bmatrix}
	e^{-i\frac{\phi}{2}}, \quad
	\kappa_{2o}=
	\begin{bmatrix}
		\alpha_{+o} & 0 \\
		0 & \alpha_{-o}
	\end{bmatrix}
	e^{i\frac{\phi}{2}},
	\label{eq:kappa12}
\end{align}
with
\begin{align}
	\alpha_{\pm o}=e^{-i\sgn\left(\Delta_{\pm o}\right)\arccos\left(E/\Delta_{\pm o}\right)}.
	\label{eq:alphapm}
\end{align}

\subsection{Normal-state scattering matrix and the energy spectrum}

The scattering of electrons and holes about the normal region is described by the scattering matrices $S_\mathrm{e}$ and $S_\mathrm{h}$, respectively. Together these matrices must obey certain symmetry relations as a result of the structure of the Hamiltonian~\eqref{eq:HBdG}. The first relation that must be satisfied as a consequence of particle-hole symmetry is
\begin{align}
	S_\mathrm{h}\left(E\right)= \mathcal{T} S_\mathrm{e}\left(-E\right) \mathcal{T}^{-1}.
	\label{eq:PHS}
\end{align}
The second relation that must be satisfied due to time-reversal invariance is
\begin{align}
	S_\mathrm{e}\left(E\right)= \sigma_2 S_\mathrm{e}^\mathsf{T}\left(E\right) \sigma_2.
	\label{eq:TRS}
\end{align}

The elements of $S_\mathrm{e}\left(E\right)$ change on the scale of $\hbar/t_\mathrm{dw}$, where $t_\mathrm{dw}$ is the dwell time in the junction. In terms of $t_\mathrm{dw}$ and $\Delta$, the short-junction limit requires $\Delta\ll\hbar / t_\mathrm{dw}$. Since we will consider energies on the scale of a few $\Delta$ or less, the energy dependence of the scattering matrix $S_\mathrm{e}\left(E\right)$ may be neglected,\cite{BeenakkerPRL1991} allowing us to take the scattering matrix at the Fermi level $S_\mathrm{e}\left(E\right)=S_\mathrm{e}\left(E=0\right)$.

The electron subblock of the scattering matrix has the general form
\begin{align}
	S_\mathrm{e}=
	\begin{bmatrix}
		r & t^\prime \\
		t & r^\prime
	\end{bmatrix}
	\label{eq:S1},
\end{align}
and, upon using time-reversal symmetry [Eq.~\eqref{eq:TRS}] in the single channel case of interest, we can parameterise\cite{BeriPRB2008}
\begin{align}
	r=\rho\openone_2, \quad r^\prime=\rho^\prime\openone_2, \quad t=\sigma_2{t^\prime}^\mathsf{T}\sigma_2=\sqrt{\tau}U,
	\label{eq:param}
\end{align}
where $\rho$ and $\rho^\prime$ are complex numbers. The transmission probability $\tau$ encodes the normal-state conductance as $G=\left(2e^2/h\right) \tau$. The spin-orbit scattering has been introduced through the $2\times 2$ unitary matrix $U=\widetilde{U}e^{i\chi}$, where $\chi$ is a real phase and $\widetilde{U}$ is an SU$(2)$ matrix which can be parameterised by Euler angles ($\theta, \omega, \eta$) via
\begin{align}
	\widetilde{U}=e^{-i\frac{\theta}{2}\sigma_3}e^{-i\frac{\omega}{2}\sigma_2}e^{-i\frac{\eta}{2}\sigma_3},
	\label{eq:uparam}
\end{align}
where the parameter $\omega$ is a measure of the degree of spin-flip scattering. Furthermore, due to the unitarity of $S_\mathrm{e}$, we have the following identity:
\begin{align}
	\rho\rho^\prime=-e^{i2\chi}\left(1-\tau\right).
	\label{eq:rhorho}
\end{align}

At this point, we are now ready to consider specific junction setups and how the Andreev energies may be obtained.

\section{Andreev bound-state spectrum}
\label{sec:AndreevBS}

The scattering processes described in the previous section lead to Andreev bound states in the normal region of the Josephson junction, where the bound states have a spectrum of energies dependent on $\phi$. These energies are the roots of the secular equation\cite{BeenakkerPRL1991}
\begin{align}
	\det\left[\openone_{4}-r_\mathrm{he}\left(E\right)S_\mathrm{e}\left(E\right)r_\mathrm{eh}\left(E\right)S_\mathrm{h}\left(E\right)\right]=0,
	\label{eq:secular1}
\end{align}
where we have introduced $\openone_{4}$ as the $4\times 4$ identity matrix. We now turn to obtaining the Andreev levels by solving the secular equation for the junction setups depicted in Figs.~\ref{fig:setups}(a)~and~\ref{fig:setups}(b).

\subsection{Junction between topological superconductors}
\label{sec:ps-ps}

Here we work towards obtaining the spectrum of Andreev bound states in the topological--topological junction depicted in Fig.~\ref{fig:setups}(a). For simplicity, we take the pairing strengths to be identical on either side of the junction, allowing us to suppress the $o$ index for convenience.

By substituting the scattering matrix~\eqref{eq:S1} and the Andreev reflection matrices~\eqref{eq:andreev0} into the secular equation~\eqref{eq:secular1} and also employing Eq.~\eqref{eq:rhorho}, the particle-hole and time-reversal symmetries in Eqs.~\eqref{eq:PHS}~and~\eqref{eq:TRS}, and the folding identity
\begin{align}
	\det
	\begin{bmatrix}
		A & B \\
		C & D
	\end{bmatrix}
	=\det\left(AD-ACA^{-1}B\right),
	\label{eq:folding}
\end{align}
the secular equation may be recast into the form
\begin{align}
	\det\left[\left(1-\tau-\Re\gamma\right)\openone_2 + \frac{\tau}{2} \left(Ye^{-i\phi}+ Y^\dagger e^{i\phi}\right) \right]=0,
	\label{eq:toptopsecular0}
\end{align}
where we have used $\gamma=\alpha_+\alpha_-$, $Y=\tilde{\kappa}_1 \widetilde{U} \tilde{\kappa}_1 \widetilde{U}^\dagger$, and $\tilde{\kappa}_1=e^{i\phi/2} \gamma^{-1/2} \kappa_1$. Since $Y \in \text{SU}(2)$, we may rewrite Eq.~\eqref{eq:toptopsecular0} as
\begin{align}
	\det\left[\left(1-\tau-\Re\gamma\right)\openone_2 + \tau \Re \left(D_Y e^{i\phi}\right) \right]=0,
	\label{eq:toptopsecular1}
\end{align}
where $D_Y$ is the diagonal matrix of eigenvalues of $Y$. We now demonstrate that the only relevant $\text{SU}(2)$ parameter in Eq.~\eqref{eq:uparam} is $\omega$. Directly substituting this parameterisation into $Y$, we find that
\begin{align}
	Y= e^{-i\frac{\theta}{2}\sigma_3}\left(\tilde{\kappa}_1 e^{-i\frac{\omega}{2}\sigma_2} \tilde{\kappa}_1 e^{i\frac{\omega}{2}\sigma_2}\right) e^{i\frac{\theta}{2}\sigma_3},
\end{align}
so that the eigenvalues contained in $D_Y$ are those of
\begin{align}
	Y^\prime = \tilde{\kappa}_1 e^{-i\frac{\omega}{2}\sigma_2} \tilde{\kappa}_1 e^{i\frac{\omega}{2}\sigma_2}
\end{align}
and hence only the parameter $\omega$ plays a role.

Since the determinant in the secular equation~\eqref{eq:toptopsecular1} is of a diagonal matrix, the secular equation amounts to either one or both of two equations being satisfied. Introducing $E_\pm=E/\Delta_\pm$ and $x_\pm=1-z_\pm\cos^2\left(\omega/2\right)$, where
\begin{align}
	z_\pm=1-E_+ E_- \pm \sqrt{1-E_+^2}\sqrt{1-E_-^2},
\end{align}
the two equations are
\begin{align}
	z_-+\tau\left[x_+\cos\left(\phi\right) \pm \sqrt{1-x_+^2}\sin\left(\phi\right) - 1\right]=0
	\label{eq:toptopsecular2}
\end{align}
and these equations together can contribute up to four solutions to the Andreev spectrum.

\subsubsection{Without $s$-wave pairing} \label{sec:3a}

If we consider the special case in which $\Delta_+=-\Delta_-$ (i.e. setting the $s$-wave admixture to zero), a compact expression may be obtained for the bound-state energies
\begin{align}
	E=\pm^\prime\Delta_+\sqrt{\tau}\cos\left(\frac{\phi\pm\omega}{2}\right),
	\label{eq:energies1}
\end{align}
where the $\pm^\prime$ has a superscript prime in order to distinguish it from the $\pm$. Equation \eqref{eq:energies1} already illustrates the effect of the spin-orbit scattering: as it is varied, it translates half of the Andreev levels to the left and the other half to the right in $E$-$\phi$ space. Furthermore, as a result of time-reversal symmetry, the branch crossings at $\phi=n\pi$, where $n$ is an integer, are protected by Kramers' theorem. Equation \eqref{eq:energies1} is a result in line with a related system in which the normal region is modelled as a tunnel barrier with a $\delta$-function potential.\cite{KwonEPJB2004}

\subsubsection{The effect of an $s$-wave pairing component} \label{sec:3c}

\begin{figure}[t]
\centering
\includegraphics[width=1\columnwidth]{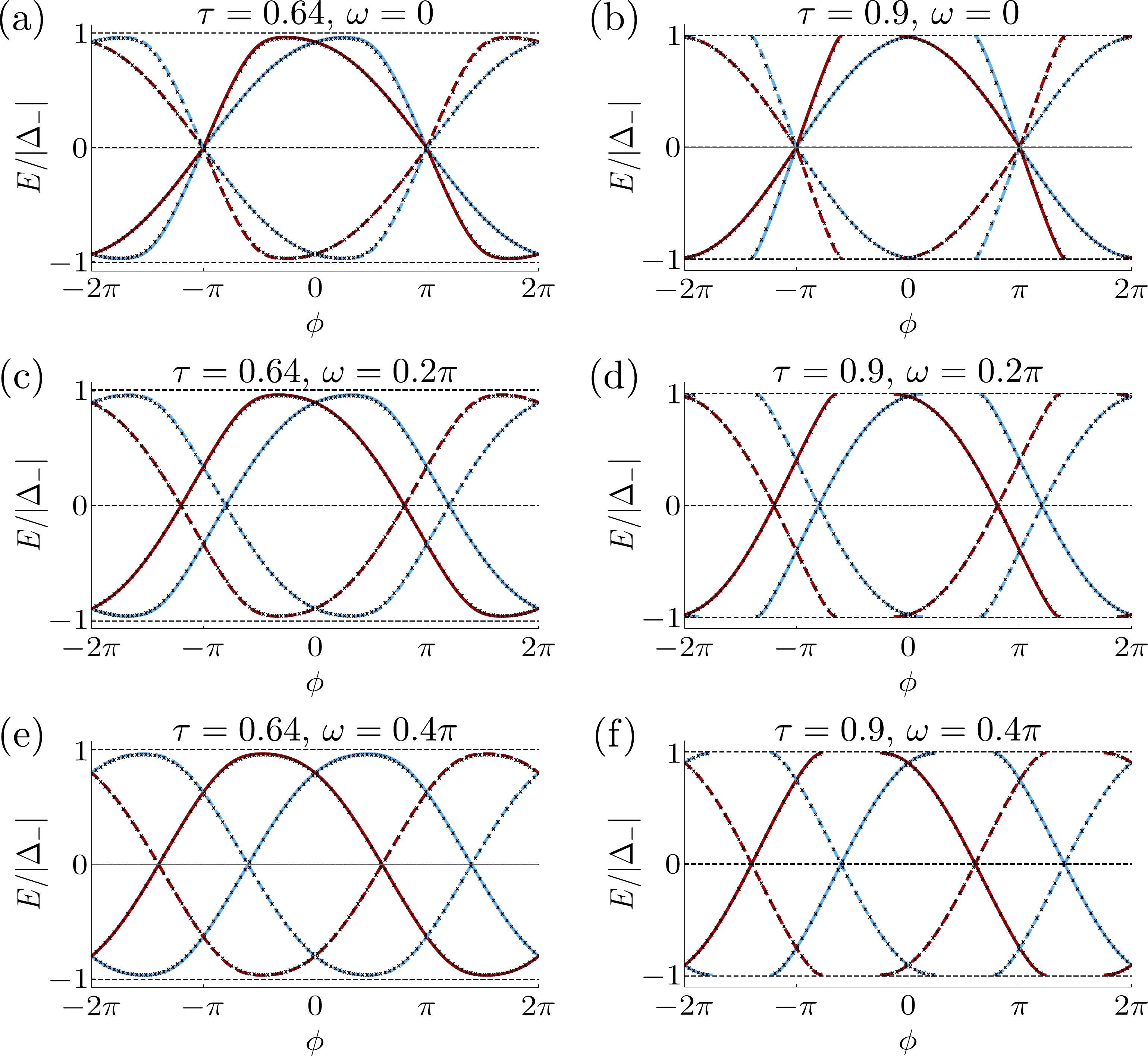}
%\captionsetup{font=small,justification=raggedright,singlelinecheck=false}
\caption{(Colour online) Andreev bound-state energies for a Josephson junction between topological superconductors with a weak $s$-wave pairing $(\Delta_s<\Delta_p)$ corresponding to $\Delta_+=-2\Delta_-$. The transmission probability $\tau$ and spin-orbit parameter $\omega$ vary throughout. The transmission probability above which bound states are lost to the continuum is $\tau_\mathrm{c}=0.75$ [see Eq.~\eqref{eq:criticalT}]. Here and in what follows, we express results in terms of the gap parameter with the smallest magnitude, and we also show data (black crosses/dashed lines) from a numerical lattice simulation (see Sec.~\ref{sec:numerics}) of the corresponding junctions. The gaps in the simulations here are $\Delta_{0\mathrm{L}}=\Delta_{0\mathrm{R}}=-0.032$ and $\Delta_{1\mathrm{L}}=\Delta_{1\mathrm{R}}=0.054$; their parameterisations in terms of $\Delta_\pm$ may be found in Eq.~\eqref{eq:LatticeEDPairings}.}
\label{fig:P+SAndreevEnergies}
\end{figure}

For the case in which $\Delta_+ \neq -\Delta_-$ (i.e. in the presence of an $s$-wave pairing admixture), the secular equation~\eqref{eq:toptopsecular1} may be solved analytically, but such an expression is cumbersome due to its size. However, it displays a number of salient and robust features shared with the purely $p$-wave case. The first of these is the crossing of the energy branches at $\phi=n\pi$ (with $n$ integer); as these are time-reversal-invariant phases, this is simply a consequence of Kramers' theorem.  A less obvious finding regards the zero crossings $\phi_n$ of the energies, defined by $E\left(\phi_n\right)=0$. From the secular equation~\eqref{eq:toptopsecular1}, we analytically extract the values of $\phi_n$ as
\begin{align}
	\phi_n=\pm\omega +\left(2n+1\right)\pi.
	\label{eq:zerocrossings}
\end{align}
The locations of these zero crossings are identical to the case of a junction without $s$-wave pairing, as in Eq.~\eqref{eq:energies1}, meaning that the inclusion of a weak $s$-wave pairing does not move or remove the zero crossings. As a result of this, the Andreev branches will remain $4\pi$ periodic in the presence of an $s$-wave pairing. (Strictly speaking, in the linear junctions we consider, the zero crossings and the consequent $4\pi$ periodicity are only approximate, neglecting the hybridisation with the Majorana--Kramers pairs at the far ends of the wires given its exponential suppression in system size. Exact zero crossings arise in a ring geometry when the only Majorana fermions in the system are in the junction.) The bound-state energies for this junction display one of the  electron-hole and time-reversal symmetry protected topological patterns proposed by Zhang and Kane\cite{ZhangKanePRB2014} in the context of anomalous topological pumps. Examples of the subgap energies are depicted in Fig.~\ref{fig:P+SAndreevEnergies}.

\begin{figure}[t]
\centering
\includegraphics[width=1\columnwidth]{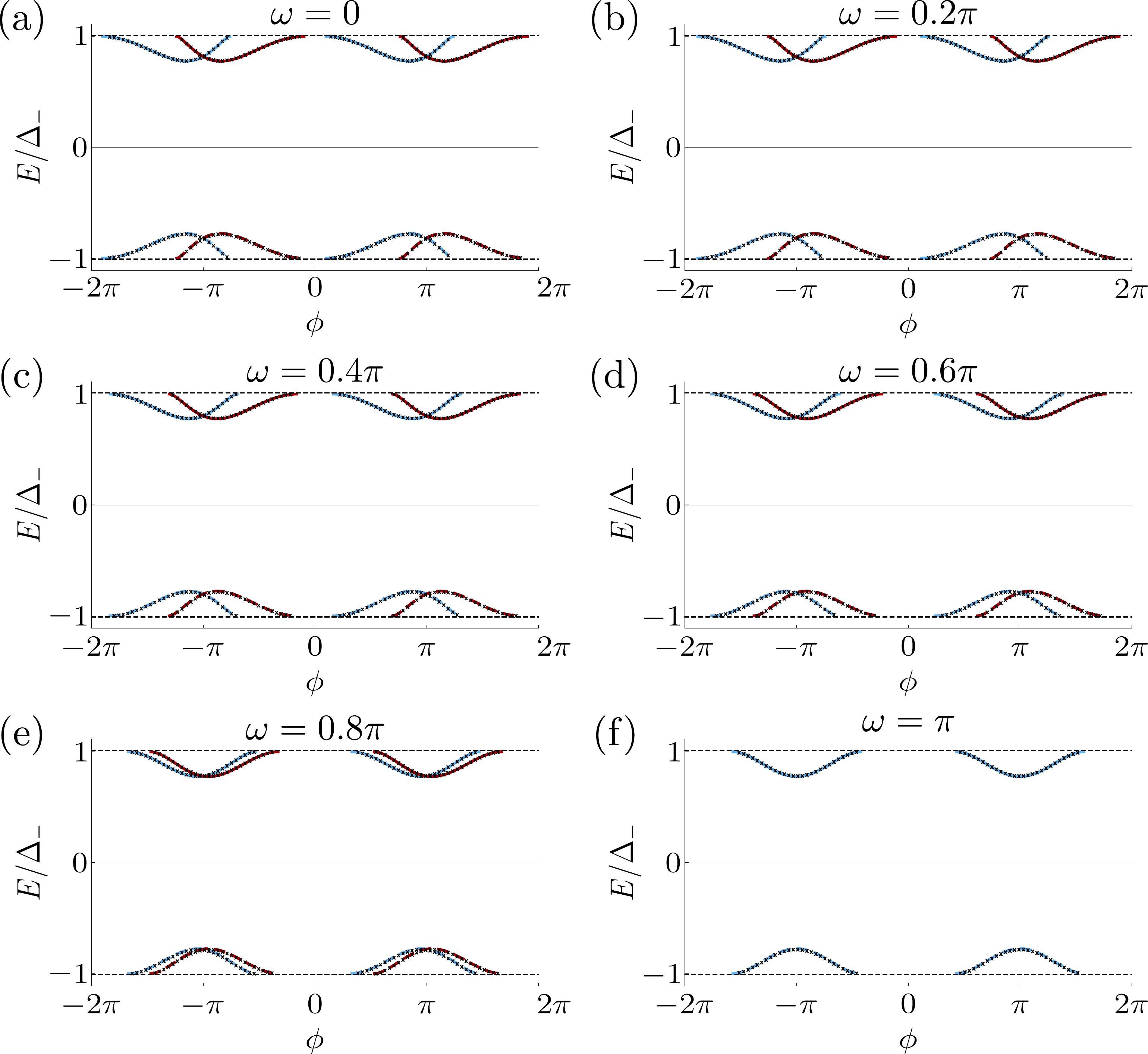}
%\captionsetup{font=small,justification=raggedright,singlelinecheck=false}
\caption{(Colour online) Andreev bound-state energies for a Josephson junction between nontopological superconductors ($\Delta_s>\Delta_p$) corresponding to $\Delta_+=2\Delta_-$. The transmission probability is $\tau=0.64$. The spin-orbit parameter $\omega$ is varied in (a)--(f), highlighting its role in splitting the levels in $\phi$-space. Levels are lost to the continuum around $\phi=0$, which is a generic feature when $\Delta_- \neq \Delta_+$. The numerical lattice data (black crosses/dashed lines) have gap parameters ${\Delta_{0\mathrm{L}}=\Delta_{0\mathrm{R}}=0.0027}$ and $\Delta_{1\mathrm{L}}=\Delta_{1\mathrm{R}}=0.018$.}
\label{fig:NonTopP+SAndreevEnergies}
\end{figure}

An important feature of the finite $s$-wave pairing admixture is the existence of an intergap energy regime ($-\Delta_-<\left|E\right|<\Delta_+$) where the subgap energies generally cannot exist. Branches that would appear to extend past $\Delta_-$ up towards $\Delta_+$ vanish after hitting the $\Delta_-$ threshold [e.g., see Fig.~\ref{fig:P+SAndreevEnergies}(b)]. As these branches approach the threshold, their derivatives are found to smoothly tend to zero. The existence of an intergap regime has ramifications for the Josephson current and is discussed in Sec.~\ref{sec:JosephsonCurrent}.

Whether there are subgap branches that reach the $\Delta_-$ threshold is dependent only on the magnitude of $\tau$ and the pairings $\Delta_\pm$. We find a critical transmission probability $\tau_\mathrm{c}$ above which branches of the Andreev spectrum are lost to the intergap regime. It has the form
\begin{align}
	\tau_\mathrm{c}=\frac{\Delta_+ - \Delta_-}{2\Delta_+}.
	\label{eq:criticalT}
\end{align}
An exception is the case of perfect transmission and a conserved $z$-component of spin (i.e. $\tau=1$ and $\omega=0$), where the $\Delta_+$ and $\Delta_-$ modes are not mixed and hence the subgap energies are not lost to the continuum.

\subsubsection{Nontopological variant}

In the case in which the $s$-wave admixture is stronger than the $p$-wave pairing (i.e. $\Delta_->0$), the superconductor is nontopological. From the secular equation~\eqref{eq:toptopsecular1}, the Andreev energies for a nontopological junction are given by the solutions to the two equations
\begin{align}
	z_++\tau\left[x_-\cos\left(\phi\right) \pm \sqrt{1-x_-^2}\sin\left(\phi\right) - 1\right]=0.
	\label{eq:nontopsecular}
\end{align}
From this equation, it may be demonstrated that, other than the exceptional case of $\tau=1$, there are no zero-energy modes for nontopological junctions.

%As the Andreev levels for a nontopological junction naturally extend up to the gap $\Delta_-$, then, provided that $\Delta_-\neq\Delta_+$, there will always be Andreev levels lost to the continuum regardless of the strength of $\tau$. In contrast to the topological case, a higher $\tau$ value actually results in a smaller region of $\phi$-space where states are lost. This is because the Andreev energies are flatter for small $\tau$ and hence do not extend very far away from the gap.

As a consequence of time-reversal symmetry, the Andreev levels must cross at phases $\phi=n\pi$. We also note that the spin-orbit parameter plays a role in splitting up the Andreev levels in $\phi$-space; this is depicted for various values of $\omega$ in Fig.~\ref{fig:NonTopP+SAndreevEnergies}. These levels correspond to the trivial pump in terms of Ref.~\onlinecite{ZhangKanePRB2014}.

\subsection{$s$-wave--topological superconductor junctions}
\label{sec:s-ps}

We now investigate the Andreev bound-state energies in the $s$-wave--topological superconductor setup, depicted in Fig.~\ref{fig:setups}(b). Specifying that the $s$-wave superconductor is to the left of the topological superconductor, the Andreev reflection matrices are as in Eqs.~\eqref{eq:andreev0}~and~\eqref{eq:alphapm}, where the gap parameters have the form
\begin{align}
	\Delta_\mathrm{\pm L}=\Delta_0 e^{-i\frac{\phi}{2}}, \quad \Delta_\mathrm{\pm R}=\Delta_\pm e^{i\frac{\phi}{2}},
\end{align}
and we also take $\alpha_{\pm\mathrm{L}}=\alpha_{0}$ and $\alpha_{\pm\mathrm{R}}=\alpha_\pm$. Now by employing the parameterisation of the scattering matrix~\eqref{eq:S1} and the folding identity~\eqref{eq:folding}, the secular equation~\eqref{eq:secular1} may be brought to the form
\begin{align}
	\det\left(\Im {\gamma_\mathrm{L}}^\frac{1}{2} \Im {\gamma_\mathrm{R}}^\frac{1}{2}\openone_2 - \frac{\tau}{2} \Re \frac{{\gamma_\mathrm{L}}^\frac{1}{2} \openone_2 - \kappa_{1\mathrm{R}} e^{-i\frac{\phi}{2}}}{{\gamma_\mathrm{R}}^\frac{1}{2}}\right)=0,
	\label{eq:nontoptopsecular1}
\end{align}
where we have introduced $\gamma_\mathrm{L}={\alpha_0}^2$ and $\gamma_\mathrm{R}=\alpha_{-} \alpha_{+}$ and the branch choice of the square root is irrelevant as long as the same choice is made for both $\gamma_\mathrm{L}$ and $\gamma_\mathrm{R}$. Note here that the secular equation is already independent of the spin-orbit scattering.  Furthermore, it may be demonstrated from this equation that the zero crossings occur when $\phi=n\pi$, where $n$ is an integer. This is precisely the behaviour one expects as for these phases the junction realises a time-reversal-invariant interface between a topological and a nontopological gapped system that must harbour a Kramers pair of Majorana zero modes. (The zero crossings, again, are strictly speaking approximate, neglecting the hybridisation with the Majorana--Kramers pairs at the far ends of the wire.)

\subsubsection{No $s$-wave admixture}

A number of interesting limiting cases exist in which the subgap energies have a compact solution. If we first consider the gap-symmetric case (i.e. $\Delta_{\pm}=\pm\Delta_0$), the Andreev bound-state energies have the analytical solution
\begin{align}
	E=\pm^\prime{\frac{\Delta_0}{\sqrt{2}}\sqrt{1 \pm \sqrt{1-\tau^2\sin^2\left(\phi\right)}}},
	\label{eq:lowandhighenergies1}
\end{align}
where two of the bound states are low energy (corresponding to the negative sign of $\pm$) and the other two are high energy (corresponding to the positive sign of $\pm$). This result is in accordance with some related models of this junction that use certain specific choices for the tunnel barrier potential.\cite{KwonEPJB2004, ElsterPRB2016}

Another interesting limit is when the conventional superconductor is strong compared to the topological superconductor (i.e. $\Delta_0 \gg \Delta_+$), where the subgap energies are
\begin{align}
	E=\pm\frac{\tau}{2-\tau} \Delta_+ \sin\left(\phi\right).
	\label{eq:strongconventional}
\end{align}
In the converse case (i.e. $\Delta_+ \gg \Delta_0$), the subgap energies are independent of the transmission probability, taking the form
\begin{align}
	E=\pm\Delta_0 \sin\left(\phi\right),
	\label{eq:strongtopological}
\end{align}
provided that $\tau>0$. The $\tau$ independence of the energies in this limit can be attributed to the fact that, for $\Delta_+ \gg \Delta_0$, on the effectively $p$-wave side, only $\alpha\left(E\rightarrow 0\right)$ is involved in the Andreev reflections, which therefore become resonant.\cite{KTLawPRL2009, Duckheim2011, WimmerNJP2011}

In the absence of an $s$-wave admixture and also where the left and right superconducting gaps differ in magnitude (i.e. $\Delta_{+} = -\Delta_- \neq \Delta_0 $), an intergap regime opens up where it is possible for Andreev levels to escape into the continuum. This result is in agreement with Ioselevich \textit{et al.}\cite{IoselevichPRB2016} and it also explains why the high-energy solutions are absent in Eqs.~\eqref{eq:strongconventional}~and~\eqref{eq:strongtopological}.

\subsubsection{Generic $s$-wave--topological junctions}

Upon the inclusion of a small $s$-wave pairing component in the topological superconductor, we find that a second intergap regime ($-\Delta_{-}<\left|E\right|<\Delta_{+}$) opens up where the Andreev bound states are able to escape into the continuum. This second intergap regime is of the same type as in the case of topological--topological superconductor junctions in Sec.~\ref{sec:3c}.

With both of these intergap regimes, there are three types of junction that are possible depending on the relative magnitudes of the superconducting gaps. In general, the intergap regime spans over the range of energies $\min\left\{\Delta_0, \left|\Delta_-\right|, \Delta_+\right\}<\left|E\right|<\max\left\{\Delta_0, \left|\Delta_-\right|, \Delta_+\right\}$.

\begin{figure}[t]
\centering
\includegraphics[width=1\columnwidth]{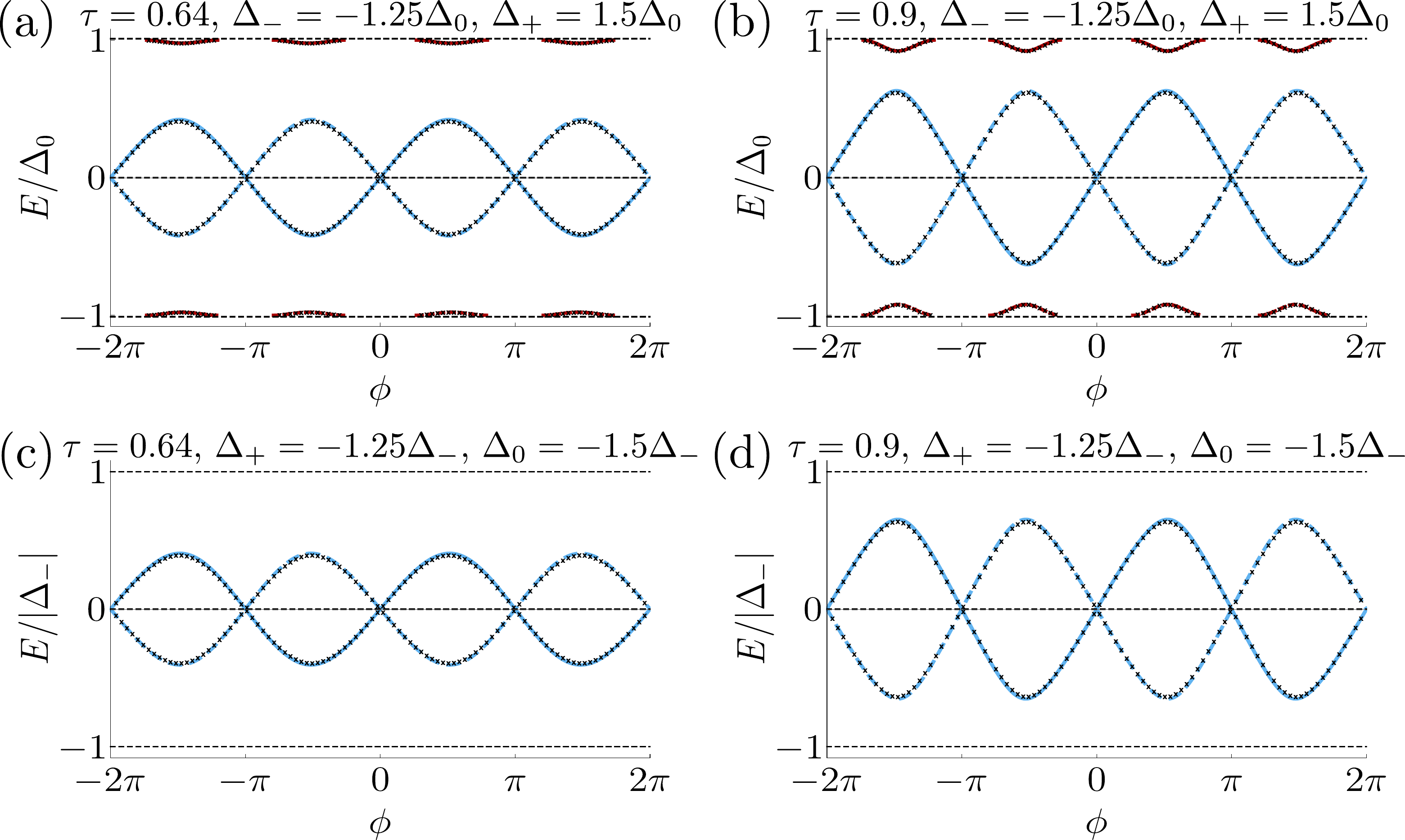}
%\captionsetup{font=small,justification=raggedright,singlelinecheck=false}
\caption{(Colour online) Andreev bound-state energies for a Josephson junction between an $s$-wave superconductor (with gap $\Delta_0$) and a topological superconductor (with gaps $\Delta_\pm$) with a finite $s$-wave pairing admixture. For (a) and (b), high-energy Andreev levels are generally present, while for (c) and (d) they are lost to the continuum for all phase differences $\phi$. The numerical lattice data (black crosses/dashed lines) uses gap parameters $\{\Delta_{0\mathrm{L}}, \Delta_{1\mathrm{L}},  \Delta_{0\mathrm{R}}, \Delta_{1\mathrm{R}}\}$ with (a)~and~(b) $\{0.02, 0, -0.065, 0.099\}$ and (c)~and~(d) $\{0.03, 0, -0.053, 0.081\}$.}
\label{fig:SandP+SAndreevEnergies}
\end{figure}

We find that the feature of a pair of high-energy solutions and a pair of low-energy solutions that arose in the gap-symmetric case remains generally true in the gap-asymmetric case with one caveat: as the high-energy solutions would appear to extend up to the largest gap of the system, high-energy subgap states are generally lost to the continuum as the intergap regime always spans from the smallest gap parameter to the largest. Illustrative examples of the bound-state spectrum are depicted in Fig.~\ref{fig:SandP+SAndreevEnergies}.

\subsubsection{Nontopological variant}

\begin{figure}[t]
\centering
\includegraphics[width=1\columnwidth]{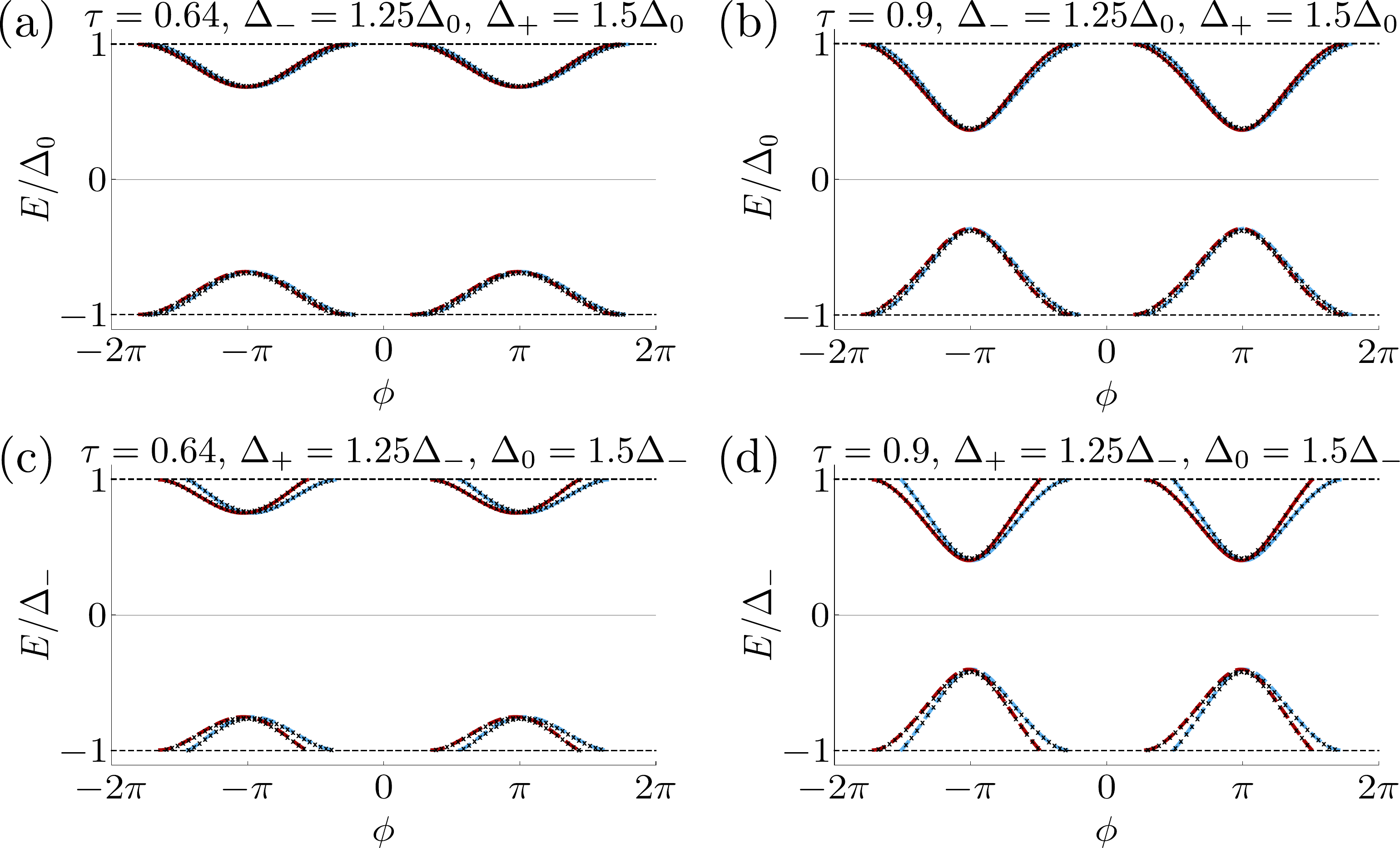}
%\captionsetup{font=small,justification=raggedright,singlelinecheck=false}
\caption{(Colour online) Andreev bound-state energies for a Josephson junction between an $s$-wave superconductor with pairing $\Delta_0$ and a nontopological superconductor with pairings $\Delta_+$ and $\Delta_->0$. The numerical lattice data (black crosses/dashed lines) uses gap parameters $\{\Delta_{0\mathrm{L}}, \Delta_{1\mathrm{L}}, \Delta_{0\mathrm{R}}, \Delta_{1\mathrm{R}}\}$ with (a)~and~(b) $\{0.02, 0, 0.021, 0.009\}$ and (c)~and~(d) $\{0.03, 0, 0.016, 0.009\}$.} 
\label{fig:nontopSandP+SAndreevEnergies}
\end{figure}

The system becomes nontopological when the $s$-wave admixture becomes greater than the $p$-wave pairing (i.e. $\Delta_->0$ in our convention). From the secular equation~\eqref{eq:nontoptopsecular1}, it follows that, in this case, Andreev levels cross zero energy only in the exceptional case $\tau=1$ at phase differences $\phi=(2n+1)\pi$ with $n$ an integer. The Majorana--Kramers pairs at each $\phi=n\pi$ are now absent. Various nontopological subgap energies are depicted in Fig.~\ref{fig:nontopSandP+SAndreevEnergies}.

\section{Josephson current}
\label{sec:JosephsonCurrent}

The Josephson current $I$ for our junctions comes from two contributions. One of these has been our focus so far: the Andreev bound-state spectrum---states which are confined to the normal region of the Josephson junction. The other current contribution originates from the continuous spectrum---states with an energy larger than the smallest of the superconducting gaps, allowing them to extend into their respective superconductors. The total current is
\begin{align}
	I=I_\mathrm{A}\left(\phi\right) + I_\mathrm{cont}\left(\phi\right),
	\label{eq:totalcurrent}
\end{align}
where $I_\mathrm{A}$ is the current due to the Andreev bound-states and $I_\mathrm{cont}$ is the current due to the continuous spectrum. In what follows, for simplicity, we will consider the zero-temperature limit and some related cases in which the Bogoliubov--de Gennes energy levels have definite occupancies. (We will comment on finite-temperature considerations in Sec.~\ref{sec:concl}.) In this case, each of the contributions can be calculated as
\begin{align}
	I_X\left(\phi\right)=\frac{2e}{\hbar}\frac{\mathrm{d}}{\mathrm{d}\phi}E_X\left(\phi\right), \qquad X = \mathrm{A}, \ \text{cont},
	\label{eq:currentcontributions}
\end{align}
where $E_X\left(\phi\right)$ is the contribution of the part $X$ of the spectrum to the total energy. We find, for both of our junction setups, that the contribution $I_\mathrm{cont}$ is zero for energies above the largest superconducting gap of the junction, while it is generally nonzero for energies that lie within an intergap regime. $I_\mathrm{A}$ is also generally nonzero for both of our topological junction setups.

The current that is measured depends on the speed at which $\phi$ changes relative to the energy and fermion parity relaxation times of the junction. In what follows, given that fermion parity relaxation often involves\cite{FuKanePRB2009} energy relaxation but not necessarily vice versa, we consider three complementary cases:  (i) when both energy and fermion parity relaxation can be neglected, (ii) with fast energy relaxation but fermion parity conservation, and (iii) when both energy and fermion parity relaxation are fast. In all cases, we use a protocol where the intended sweep is preceded by some period of slow sweep in regime (iii), which ensures definite Bogoliubov--de Gennes level occupancies. In addition to this, we assume the sweep  speeds described in all cases are slow enough that unintended Landau-Zener tunnelling between branches or to the continuum is avoided.

Potential mechanisms for these relaxation processes are phonon or photon coupling in the case when parity is conserved, and quasiparticle poisoning (e.g., from bulk localised states\cite{FuKanePRB2009}) when it is not. In terms of typical timescales for these processes, we assume that, depending on the relaxation regime in which one works, the time taken to traverse a $4\pi$ period of the junction is either much faster or much slower than the relevant relaxation timescales. A recent experiment\cite{WoerkomNature2015} on a Majorana-related mesoscopic superconductor system has been conducted working in the regime where the dominant relaxation respects fermion parity.

\subsection{Junctions between topological superconductors}

The contribution to the Josephson current by the Andreev levels is dependent on the fermion parity of the junction. As the Andreev levels come in particle-hole pairs, only half of them may be occupied at a given moment (due to the redundancy $c^\dagger_E=c_{-E}^{}$ of the corresponding fermion operators). If we associate one pair to have energy $\pm E_1\left(\phi\right)$ and the other pair to have energy $\pm E_2\left(\phi\right)$, where the Andreev levels are labelled as in Fig.~\ref{fig:ParityDemo}(a), then the total energy from the Andreev levels is\cite{ChtchelkatchevPRL2003}
\begin{align}
	E_\mathrm{A} \left(\phi\right)=-\frac{1}{2}\left[\left(-1\right)^{n_1} E_1\left(\phi\right) + \left(-1\right)^{n_2}E_2\left(\phi\right)\right],
	\label{eq:parityenergies}
\end{align}
where $p=n_1+n_2 \pmod{2}$ is the fermion parity of the junction. In using the quantum numbers $n_j$, we neglect the hybridisation with the Majorana--Kramers pairs at the far ends of the wires, as this is exponentially suppressed with system size. (This approximation influences only the parity-conserving cases discussed below, where it amounts to assuming that $\phi$ changes quickly enough so that Landau-Zener tunnelling occurs with probability unity across the exponentially small splittings that the zero crossings approximate.) The concrete choice of $n_j$, as $\phi$ is varied, will depend on the junction's interaction with its environment, specifically on whether the junction is able to relax to the ground state (potentially subject to a parity constraint). We will separately discuss each of these cases in what follows.

\subsubsection{Subgap current in the absence of relaxation}

We consider the possible energies and currents in the absence of energy relaxation and, additionally for Landau-Zener tunnelling to the continuum to be in principle avoidable, with no levels escaping into the continuum. We have four branches of energy for the four values of $\left\{n_1, n_2\right\}$, each of which is $4\pi$ periodic in $\phi$. This is illustrated in Fig.~\ref{fig:ParityDemo}(b). The consequent current contribution is also $4\pi$ periodic.

\subsubsection{Subgap current in the presence of energy relaxation}
\label{sec:psstrongrelaxation}

\begin{figure}[t]
\centering
\includegraphics[width=1\columnwidth]{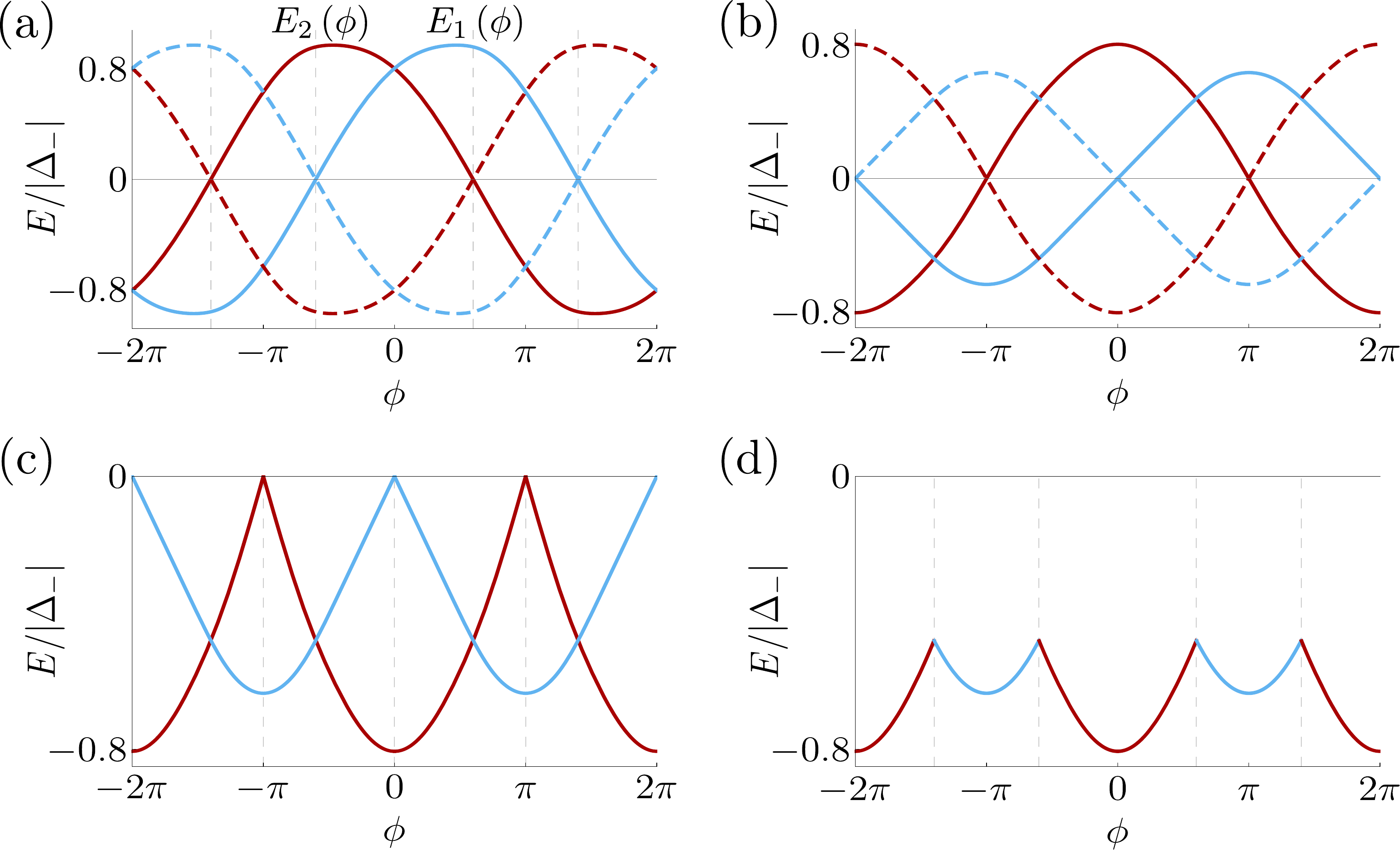}
%\captionsetup{font=small,justification=raggedright,singlelinecheck=false}
\caption{(Colour online) The subgap energies for a Josephson junction between topological superconductors. The superconductor pairing is $\Delta_+=-2\Delta_-$, while the transmission probability and spin-orbit parameter are $\tau=0.64$ and $\omega=0.4\pi$. In (a) the blue and red curves represent $E_1\left(\phi\right)$ and $E_2\left(\phi\right)$, respectively [as in Eq.~\eqref{eq:parityenergies}], while the dashed lines signify their particle-hole partners. In (b)--(d), the blue lines are odd-parity, while the red lines are even-parity energies. Fermion parity is conserved in (b) and (c), but not in (d). Energy relaxation is absent in (b), but not in (c) and (d). Numerical lattice data have been omitted, as the agreement has already been demonstrated to be excellent.}
\label{fig:ParityDemo}
\end{figure}

We now consider the subgap energy and its contribution to the current when the energy relaxation is much faster than the sweep speed of $\phi$. This amounts to choosing the minimum energy $E_\mathrm{A}\left(\phi\right)$ within a given parity sector, resulting in branch switches at the locations $\phi=\left(p+2n+1\right)\pi$. These locations correspond to those of finite-energy Andreev branch crossings, which are where energy-minimising Andreev branch occupancy switches can occur without a change in fermion parity.  Examples are depicted in Fig.~\ref{fig:ParityDemo}(c). The corresponding current contribution is $2\pi$ periodic and has jumps (``current switches") at the branch switch locations.

\subsubsection{Subgap current without fermion parity conservation.}
\label{sec:sxpSubgap}

Finally, we consider the case in which the junction relaxes to its ground state without conserving fermion parity. In this case, the junction takes the minimum energy regardless of fermion parity, resulting in energy branch switches at the locations specified by Eq.~\eqref{eq:zerocrossings}. Underlying the switches are now the zero-energy Andreev branch crossings, which, via changing the fermion parity, ensure that all of the negative energy Andreev levels are occupied for all $\phi$.  Such switches are indicative of the topological pattern of subgap branches of Ref.~\onlinecite{ZhangKanePRB2014}. Examples are shown in Fig.~\ref{fig:ParityDemo}(d). The energies and the corresponding current contribution are $2\pi$ periodic. The branch switches, again, are accompanied by switches in the current contribution. These current switches are the class DIII analogues of the switches in class D systems, discussed in Ref.~\onlinecite{MarraPRB16}.

\subsubsection{Continuous contributions to the Josephson current}

In Sec.~\ref{sec:ps-ps}, we identified the energy regime ${\left|E\right|>\left|\Delta_-\right|}$ as where the spectrum is continuous. The Josephson current due to this spectrum will depend on the density of states $\rho$, whose structure is dependent on whether the energy is in the intergap regime ${\left|\Delta_-\right|<\left|E\right|<\Delta_+}$ or the above gap regime $\left|E\right|>\Delta_+$. Such a supercurrent may be calculated according to Eq.~\eqref{eq:currentcontributions} and noting that the contribution of the filled negative energy Bogoliubov--de Gennes energy levels is $\frac{1}{2}\sum_{E_j<0}E_j$,\cite{ChtchelkatchevPRL2003}
\begin{align}
	I_{\mathrm{cont}}=\frac{e}{\hbar} \int_{-\infty}^{-\left|\Delta_-\right|} \mathop{\mathrm{d}E} E \frac{\partial\rho}{\partial\phi},
	\label{eq:contcontrib1}
\end{align}
where $\rho$ may be expressed in terms of the total scattering matrix of the junction $S_\mathrm{SNS}$,\cite{AkkermansPRL1991}
\begin{align}
	\rho=\frac{1}{2\pi i}\frac{\partial}{\partial E} \ln\left[\det\left(S_\mathrm{SNS}\right)\right] + \mathrm{constant}.
	\label{eq:DoS1}
\end{align}
The matrix $S_\mathrm{SNS}$ has the general form
\begin{align}
	S_\mathrm{SNS}=\hat{R}+\hat{T}^\prime\left(\openone-S_\mathrm{N} \hat{R}^\prime\right)^{-1} S_\mathrm{N} \hat{T},
	\label{eq:skull}
\end{align}
with $\hat{R}$ and $\hat{T}$ describing reflection and transmission for modes incoming from within the superconductor at the SN interface, $\hat{R}^\prime$ and $\hat{T}^\prime$ describing reflection and transmission for modes incoming from the normal region at the SN interface, and $S_\mathrm{N}=\diag\left(S_\mathrm{e}, S_\mathrm{h}\right)$ describing scattering off the normal region.

\begin{figure}[t]
\centering
\includegraphics[width=1\columnwidth]{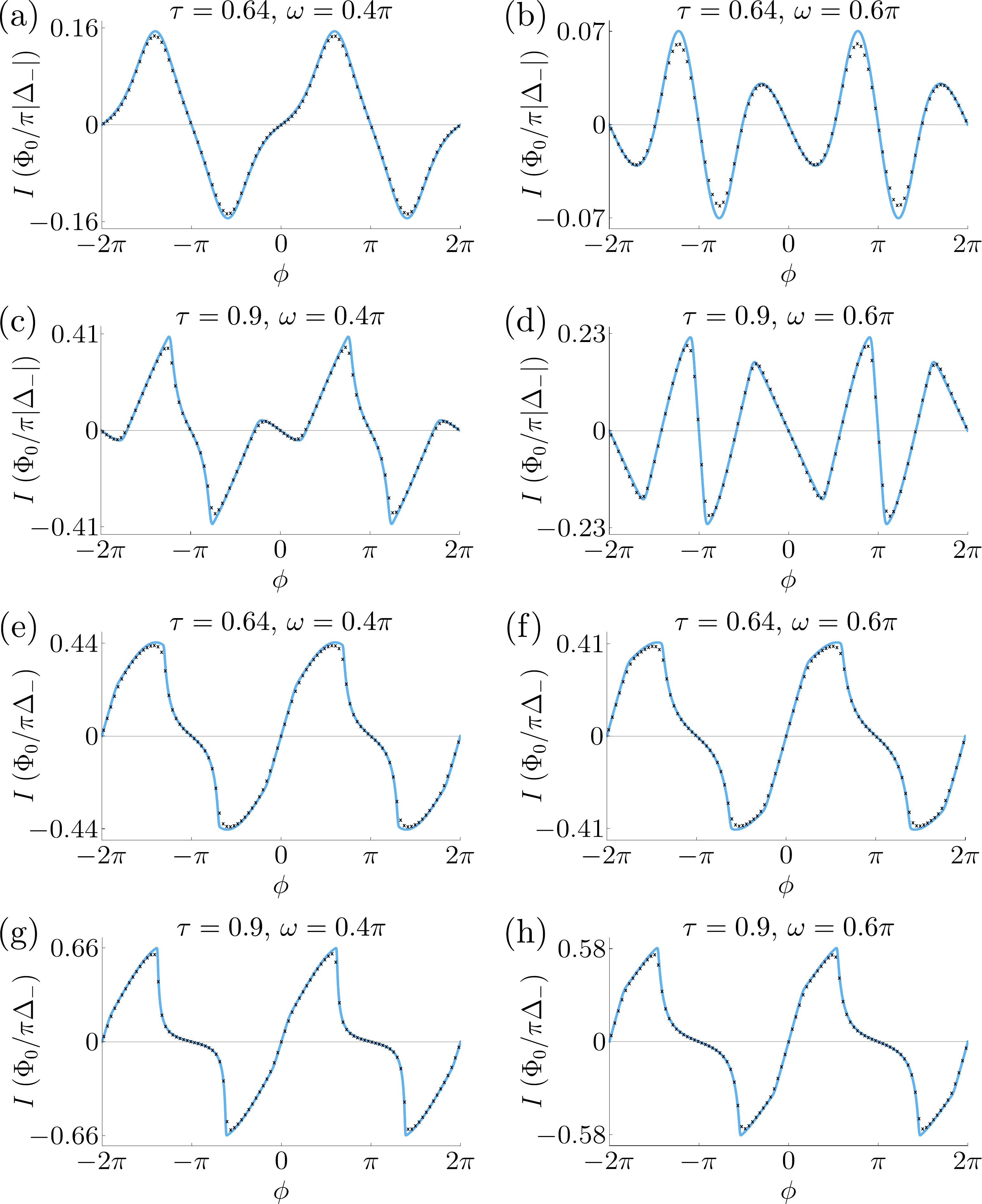}
%\captionsetup{font=small,justification=raggedright,singlelinecheck=false}
\caption{(Colour online) (a)--(d) Continuous contributions to the supercurrent for a Josephson junction between topological superconductors with gap parameter $\Delta_+=-2\Delta_-$, and (e)--(h) its nontopological counterpart with gap parameter $\Delta_+=2\Delta_-$. For topological junctions, the critical transmission above which subgap states escape into the continuum is $\tau_\mathrm{c}=0.75$ [see Eq.~\eqref{eq:criticalT}]. The numerical lattice data (black crosses) have gap parameters $\Delta_{0\mathrm{L}}=\Delta_{0\mathrm{R}}=-0.032$ and $\Delta_{1\mathrm{L}}=\Delta_{1\mathrm{R}}=0.054$ for the topological junctions, and $\Delta_{0\mathrm{L}}=\Delta_{0\mathrm{R}}=0.0027$ and $\Delta_{1\mathrm{L}}=\Delta_{1\mathrm{R}}=0.018$ for the nontopological junctions.}
\label{fig:intergapcurrents1}
\end{figure}

\textit{(a) Above the gaps.} In the above the gap regime ($\left|E\right|>\Delta_+$), we demonstrate that there are no continuous contributions to the Josephson current. The matrix $\hat{R}^\prime$ has the general form
\begin{align}
	\hat{R}^\prime=
	\begin{bmatrix}
		0 & r_\mathrm{eh} \\
		r_\mathrm{he} & 0
	\end{bmatrix},
	\label{eq:Rprime}
\end{align}
where $r_\text{eh}$ and $r_\text{he}$ are the $4 \times 4$ matrices from Eq.~\eqref{eq:andreev0} evaluated for $\left|E\right|>\Delta_+$. The other three scattering matrices are related via
\begin{align}
	\hat{R}=-\sigma_1 \hat{R}^\prime \sigma_1, \quad \hat{T}^\prime = \sqrt{\openone_8 +\hat{R}\hat{R}^\prime}, \quad \hat{T}=\sigma_1 \hat{T}^\prime \sigma_1.
	\label{eq:scatteringrelations}
\end{align}
By using the relations in Eq.~\eqref{eq:scatteringrelations} in the expression for $S_\mathrm{SNS}$~\eqref{eq:skull}, it may be shown, with some algebra, that
\begin{align}
	S_\mathrm{SNS}=\hat{T}^\prime\left(\openone_8-S_\mathrm{N}\hat{R}^\prime\right)^{-1}\left(S_\mathrm{N}\sigma_1-\sigma_1\hat{R}^\prime\right){\hat{T}}^{\prime^{-1}}\sigma_1.
	\label{eq:sSNS1}
\end{align}
Then, using that $\alpha_\pm$ are real for energies above the gaps and substituting Eq.~\eqref{eq:sSNS1} into the density of states~\eqref{eq:DoS1}, we find that
\begin{align}
	\rho=-\frac{1}{\pi}\frac{\partial}{\partial E} \Im \ln \det \left(\openone_4+aa^*\right) + \mathrm{constant},
	\label{eq:DoS2}
\end{align}
where	$a=S_\mathrm{e}^* \sigma_2  r_\mathrm{he}$. It follows that $\det\left(\openone_4+aa^*\right)$ is real and hence the density of states~\eqref{eq:DoS2} is constant. Therefore, there is no contribution to the Josephson current in the above the gap regime.

\textit{(b) Between the gaps.} For energies in the intergap regime ($\left|\Delta_-\right|<\left|E\right|<\Delta_+$) we show that there are nonvanishing contributions to the Josephson current. The matrix $\hat{R}^\prime$ has the same form as in Eq.~\eqref{eq:Rprime}; however, the rest of the matrices that describe scattering at the SN interface have reduced dimension since the $\Delta_+$ modes cannot be transmitted into the superconducting leads. As a result, $S_\mathrm{SNS}$ is a $4 \times 4$ matrix.

The continuous contributions to the Josephson current are obtained numerically by substituting Eq.~\eqref{eq:skull} into Eq.~\eqref{eq:contcontrib1}. The currents are found to be $2\pi$ periodic and their magnitude increases with the transmission probability. For small values of $\tau$, the contributions resemble sinusoidal functions, becoming increasingly nonsinusoidal as $\tau$ is increased; this must occur as the continuous contributions compensate for subgap levels that escape into the continuum above $\tau_\mathrm{c}$. Examples of continuous contributions to the current (measured in terms of the flux quantum $\Phi_0=h/2e$) are depicted in Fig.~\ref{fig:intergapcurrents1}(a)--(d) for the topological case and in Fig.~\ref{fig:intergapcurrents1}(e)--(h) for the nontopological case.

\subsubsection{Total Josephson current}

\begin{figure}[t]
\centering
\includegraphics[width=1\columnwidth]{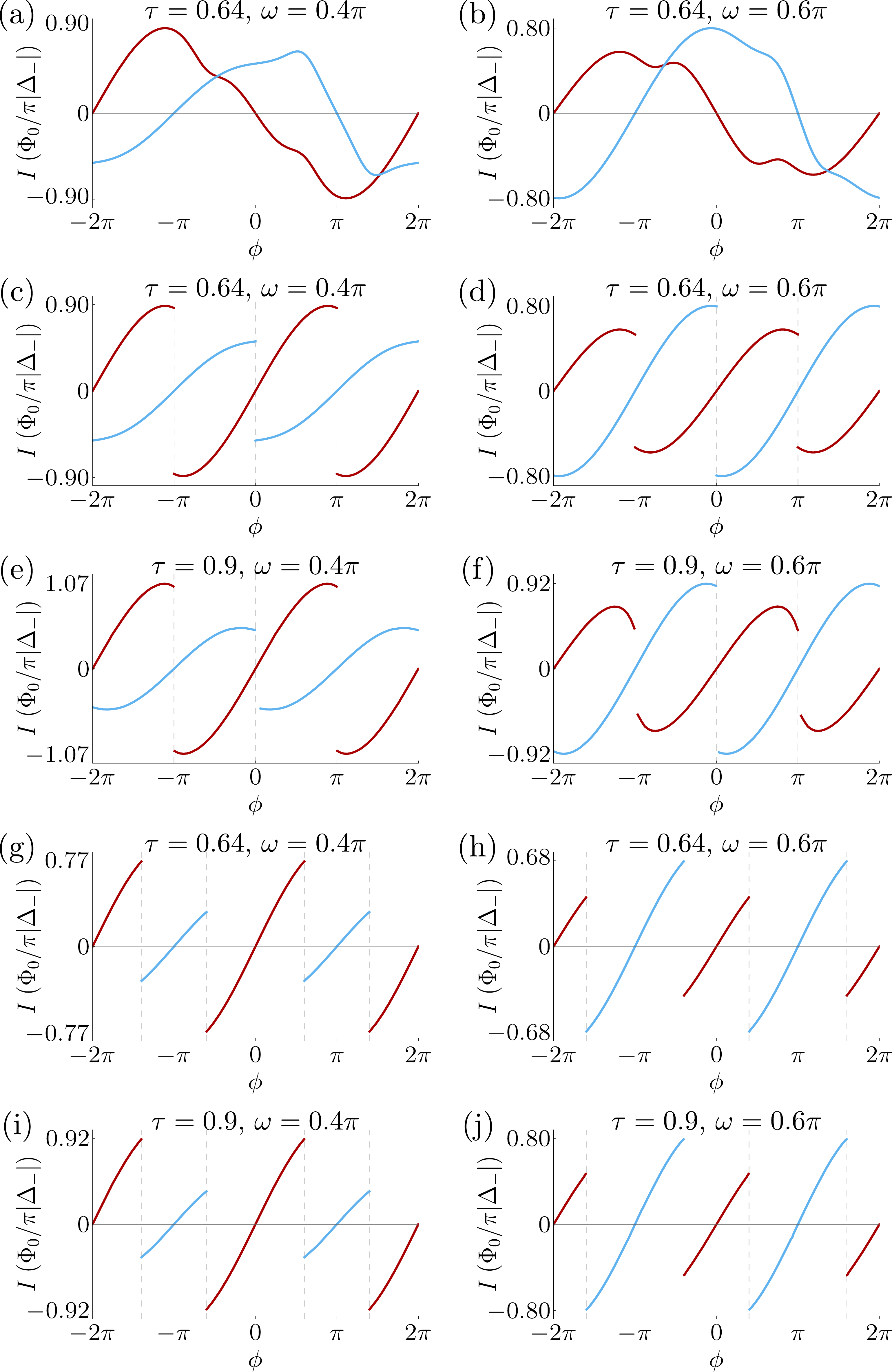}
%\captionsetup{font=small,justification=raggedright,singlelinecheck=false}
\caption{(Colour online) The total supercurrent for topological--topological junctions with gap parameter \mbox{$\Delta_+=-2\Delta_-$}. The blue and red curves represent odd- and even-parity currents. For (a)~and~(b) there is no relaxation, for (c)--(f) only energy relaxation is present, and for (g)--(j) fermion parity relaxation is also allowed. Panels (a) and (b) show one of the odd and one of the even-parity currents (not displaying the complementary two with opposite subgap contributions). Numerical lattice data have been omitted since the agreement with subgap energies and continuous current contributions has already been shown to be excellent.}
\label{fig:p+sTotalCurrents}
\end{figure}

The total current is calculated by combining the subgap and continuous contributions, as in Eq.~\eqref{eq:totalcurrent}. In the case without energy relaxation, as in the subgap case, we focus on junctions where Andreev levels do not escape into the continuum. However, in the presence of relaxation our discussion includes also the case of escaping levels. Examples are depicted in Fig.~\ref{fig:p+sTotalCurrents} for topological junctions and in Fig.~\ref{fig:nontopSandPandSCurrentVaryingT}(a) for nontopological junctions.

Generally, we find the continuous contributions to be significant relative to the subgap current when the gap asymmetry between $\Delta_+$ and $\left|\Delta_-\right|$ is appreciable. The continuous contributions are typically most significant when there are subgap states that escape into the continuum, as then the continuous contributions must account for the missing subgap levels.

There are some distinguishing features between topological and nontopological junctions. One such feature is the fractional Josephson effect: when the phase is swept much faster than the relaxation time of the junction (but not so fast that Landau-Zener tunnelling between branches or to the continuous spectrum occurs), the supercurrent of a topological--topological junction will be $4\pi$ periodic [see Figs.~\ref{fig:p+sTotalCurrents}(a)~and~\ref{fig:p+sTotalCurrents}(b)], twice that of the conventional Josephson effect. This $4\pi$ periodicity, however, is sensitive to energy relaxation, which motivates the need for a topological signature in junctions where energy relaxation takes place.

A distinctive feature appears in the case with energy relaxation: in topological Josephson junctions, the Josephson current is $2\pi$ periodic and displays a current switch at $\phi =\left(p+2n+1\right)\pi$ with $p=0,  1$ the fermion parity of the junction and $n$ an integer [see Figs.~\ref{fig:p+sTotalCurrents}(c)--(f)]. We have investigated the analogous situation in the nontopological case in which, due to the absence of Andreev level zero crossings and the use of the preparation protocol in Sec.~\ref{sec:JosephsonCurrent}, only the even parity-ground state arises and the corresponding current exhibits switchlike behaviour at $\phi=\left(2n+1\right)\pi$ when $\tau$ is close to unity [see Fig.~\ref{fig:nontopSandPandSCurrentVaryingT}(a)]. With regard to establishing switches as topological, it is most judicious to work with junctions having an intermediate value of $\tau$ in order to suppress nontopological switches, while still realising Josephson currents of an appreciable magnitude.

When both energy and fermion parity are allowed to relax, further qualitative distinctive features are found to emerge. A single pair of switches occurs in the interval $-\pi<\phi\leq\pi$, with the switch locations given by the zero crossings in Eq.~\eqref{eq:zerocrossings}. As these switches occur symmetrically about $\phi=n\pi$, junctions with a value of $\omega$ sufficiently (i.e. determined by the resolution of $\phi$) far enough away from $n\pi$ have topological switches that are distinct compared to possible nontopological counterparts. In the case in which $\omega$ is near $n\pi$, switches may be distinguished from their nontopological variants provided that $\tau$ is not too close to unity, as then potential nontopological switchlike false positives do not arise [see Fig.~\ref{fig:nontopSandPandSCurrentVaryingT}(a)].  Besides being a topological signature (as explained in Sec.~\ref{sec:sxpSubgap}), identifying the location of these switches provides a way to measure the value $\omega$ for the junction. Lastly, an important characteristic of these topological switches is that the magnitude of the current is not necessarily the same after a switch. Examples of these switches are depicted in Fig.~\ref{fig:p+sTotalCurrents}(g)--(j).

\subsection{$s$-wave--topological superconductor junctions}

For $s$-wave--topological superconductor junctions, due to the high-energy and low-energy subgap branches being generically (i.e. for $\tau$ away from unity) well separated and using the protocol outlined in Sec.~\ref{sec:JosephsonCurrent}, the cases without relaxation and with energy relaxation only coincide: in terms of the subgap current, they both correspond to always taking the positive high-energy subgap level to be empty. Therefore, it is sufficient to consider the cases distinguished by the (non)conservation of fermion parity.

\subsubsection{Subgap current with conserved fermion parity}

With a conserved fermion parity, we include the contribution of one of the low-energy branches where the choice of branch depends on the fermion parity of the junction. The resultant current has $2\pi$ periodicity.

The nature of the low-energy and high-energy contributions to the Josephson current can be understood in terms of symmetries of the system. Time-reversal symmetry relates an energy at $\phi$ to one of equal value at $-\phi$. As a result of this and the $2\pi$ periodicity of the spectrum, the high-energy Andreev branches are even in $\phi$ about the time-reversal-invariant phases $\phi=n\pi$, and the corresponding current contribution is odd, vanishing at these phases. For the low-energy branches, applying time reversal brings the energy to the opposite branch related to the initial one by particle-hole symmetry. This implies that time reversal flips fermion parity, signifying the anomalous time-reversal properties of Majorana--Kramers doublets.\cite{QiPRL2009} At the same time, this also means that the low-energy branches are odd in $\phi$ about $\phi=n\pi$, and the corresponding contribution to the current is even with finite currents at $\phi=n\pi$. Moreover, as a consequence of particle-hole symmetry, each parity sector of the low-energy Andreev levels contributes oppositely to the current.

\subsubsection{Subgap current with nonconserved fermion parity}

For a nonconserved fermion parity, the energy of the junction is always minimised and branch switches occur at $\phi=n\pi$ ($n$ is an integer), where the low-energy branches cross each other at zero energy due to the emergence of a Kramers pair of Majorana zero modes. This leads to switches in the subgap Josephson current contribution, in agreement with the work of Chung \textit{et al.}\cite{ChungPRB2013} in the tunnelling limit. As a result of the vanishing high-energy contribution at $\phi=n\pi$, these current switches are entirely due to the low-energy Andreev levels. Moreover, at the switches the sign of the current is flipped while its magnitude is preserved, which is a feature dictated by the time-reversal properties of the spectrum: the contributions just after the switch are the time-reverse of those just before the switch. The switch itself, in this sweep regime, is indicative of a flip in the fermion parity. Taken together, these magnitude-preserving sign switches therefore indicate the anomalous time-reversal properties of Majorana--Kramers pairs. As we will see below, this feature is left intact when continuous contributions are taken into consideration.

\subsubsection{Continuous contributions to the Josephson current}

The Josephson current due to the continuous contributions is calculated by substituting the total scattering matrix $S_\mathrm{SNS}$, which has the same general form as in Eq.~\eqref{eq:skull}, into the density of states~\eqref{eq:DoS1}, and finally into Eq.~\eqref{eq:contcontrib1}. We find that the continuous contributions are generally nonzero, but only when the energies lie in an intergap regime. Moreover, as with the subgap contributions, the continuous contributions are independent of the spin-orbit scattering.

\textit{(a) Above the gaps.} The calculation of the contribution to the Josephson current here is similar to the topological--topological case. The Andreev reflection matrix here assumes the same general form as in Eq.~\eqref{eq:Rprime} and the relations~$\eqref{eq:scatteringrelations}$ also hold. As a result of this, $S_\mathrm{SNS}$ takes the same form as in Eq.~\eqref{eq:sSNS1} and thus an analogue to Eq.~\eqref{eq:DoS2} can be derived (i.e. involving a determinant that is real). Therefore, the contribution to the Josephson current due to energies above the gaps is zero.

\begin{figure}[t]
\centering
\includegraphics[width=1\columnwidth]{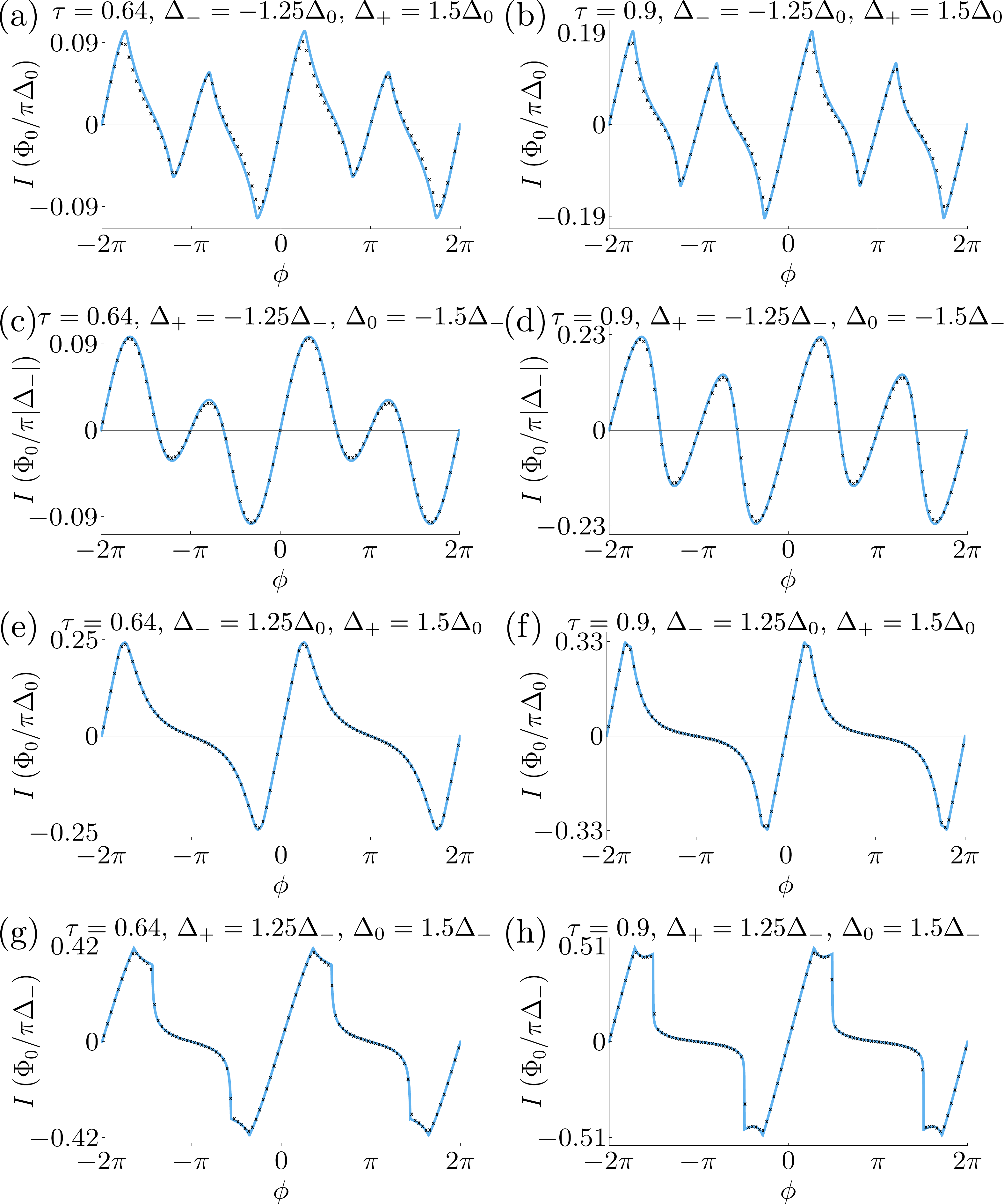}
%\captionsetup{font=small,justification=raggedright,singlelinecheck=false}
\caption{(Colour online) (a)--(d) Continuous intergap contributions to the supercurrent for an $s$-wave--topological Josephson junction and (e)--(h) its nontopological counterpart. The numerical lattice data (black crosses) have gap parameters $\{\Delta_{0\mathrm{L}}, \Delta_{1\mathrm{L}}, \Delta_{0\mathrm{R}}, \Delta_{1\mathrm{R}}\}$ with (a)~and~(b) $\{0.02, 0, -0.065, 0.099\}$, (c)~and~(d) $\{0.03, 0, -0.053, 0.081\}$, (e)~and~(f) $\{0.02, 0, 0.021, 0.009\}$, and (g)~and~(h) $\{0.03, 0, 0.016, 0.009\}$.}
\label{fig:SandPandSContContribs}
\end{figure}

\textit{(b) Between the gaps.}  As mentioned in Sec.~\ref{sec:s-ps}, there are a number of intergap energy regimes that arise. The specific form of $S_\mathrm{SNS}$ will change depending on which intergap energy window one works in (as this controls what is allowed to extend indefinitely into the superconductors) and the relative magnitudes of the gap parameters. If we specify the relevant gap parameters $\Delta=\left\{-\left|\Delta_{-}\right|, -\Delta_+, -\Delta_0\right\}$, then the intergap contributions are calculated in an analogous fashion to Eq.~\eqref{eq:contcontrib1} by the integral
\begin{align}
	I_{\mathrm{cont}}=\frac{e}{\hbar} \int_{\min\left(\Delta\right)}^{\max\left(\Delta\right)} \mathop{\mathrm{d}E} E \frac{\partial\rho}{\partial\phi}.
\end{align}
This integral is calculated numerically in the same way as with topological--topological junctions. The continuous contributions for various gap parameters and transmission probabilities can be found for the topological case in Figs.~\ref{fig:SandPandSContContribs}(a)--(d) and for the nontopological case in Figs.~\ref{fig:SandPandSContContribs}(e)--(h).

The continuous contributions are zero at $\phi=n\pi$ (with $n$ integer) as energy levels in the continuum must be even in terms of $\phi$ about the phases $\phi=n\pi$ as a result of time-reversal symmetry. Therefore, the contribution to the total current at these phases is entirely due to the low-energy Andreev levels, implying that the highlighted key features in the preceding section are not obscured.

\begin{figure}[t]
\centering
\includegraphics[width=1\columnwidth]{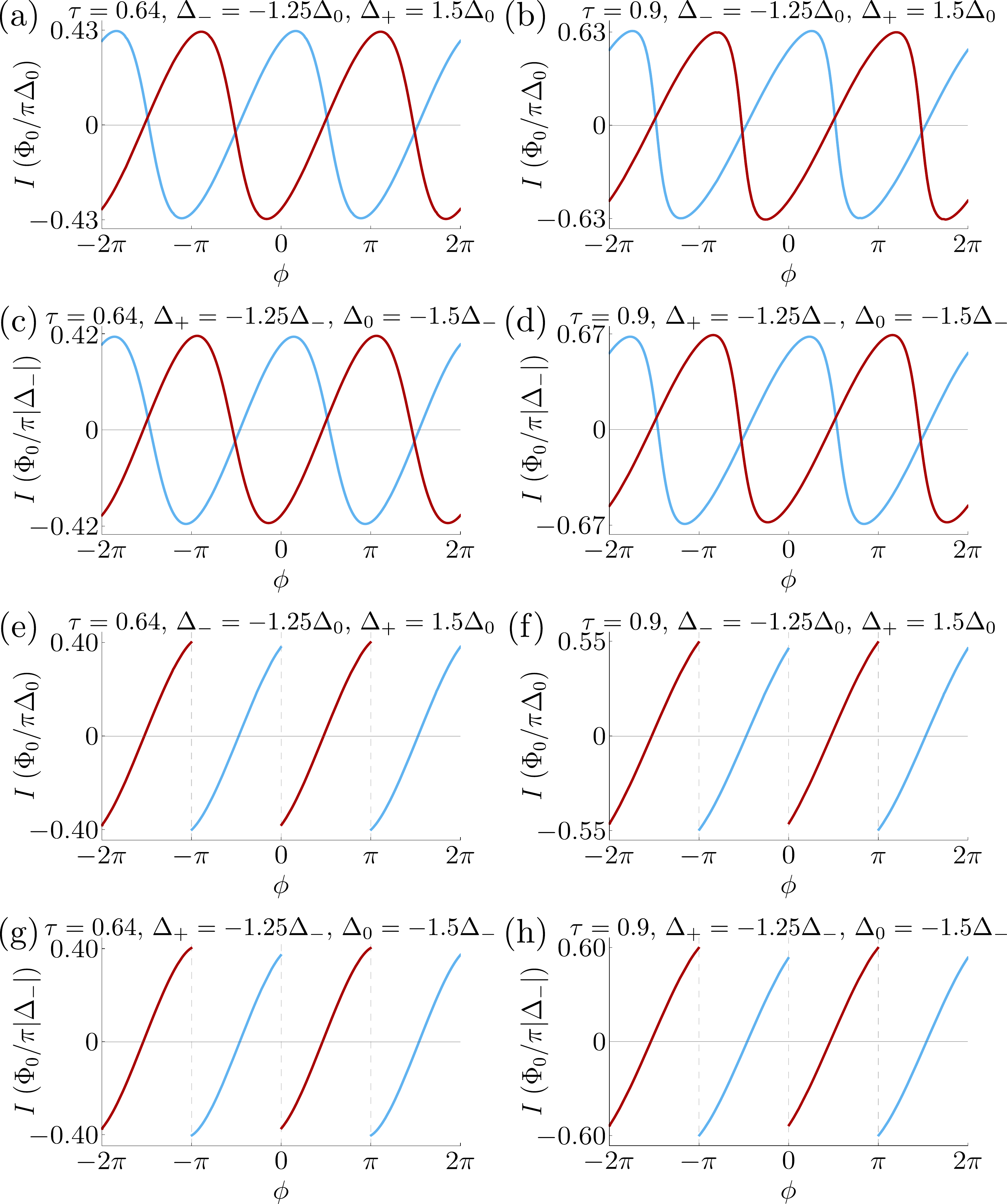}
%\captionsetup{font=small,justification=raggedright,singlelinecheck=false}
\caption{(Colour online) The total supercurrent of the $s$-wave--topological junction. The blue and red curves represent odd and even parity currents. For (a)--(d) local fermion parity is conserved, while for (e)--(h) it is violated via relaxation, resulting in current switches at phase differences $\phi=n\pi$ with $n$ an integer. Numerical lattice data have been omitted here since the agreement with subgap energies and continuous contributions to the current has already been shown to be excellent.}
\label{fig:SandPandSTotalCurrent}
\end{figure}

\subsubsection{Total Josephson current}

\begin{figure}[t]
\centering
\includegraphics[width=1\columnwidth]{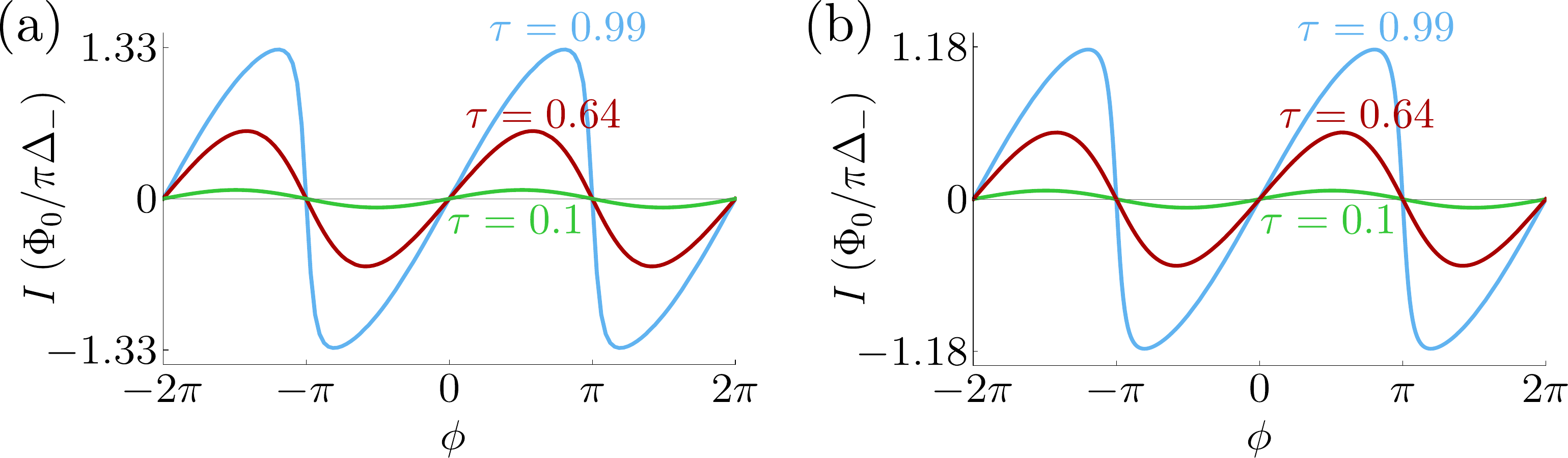}
%\captionsetup{font=small,justification=raggedright,singlelinecheck=false}
\caption{(Colour online) The total supercurrent for the nontopological variants of time-reversal-invariant Josephson junctions, where we see the emergence of switchlike features at phase differences $\phi=\pm\pi$ as the transmission probability $\tau$ approaches unity. Displayed in (a) is the total supercurrent for a nontopological--nontopological junction with spin-orbit parameter $\omega=0.4\pi$ and superconducting gaps $\Delta_+=2\Delta_-$, and in (b) the total supercurrent for an $s$-wave--nontopological junction with superconducting gaps $\Delta_+=1.25\Delta_-$ and ${\Delta_0=1.5\Delta_-}$. Lattice numerics are omitted here, as the agreement with subgap energies and continuous contributions to the current has already been shown to be excellent.}
\label{fig:nontopSandPandSCurrentVaryingT}
\end{figure}

Following Eq.~\eqref{eq:totalcurrent}, the total current is the combination of both the subgap and continuous contributions. We consider the two cases which differ by the conservation of fermion parity. Examples are showcased in Fig.~\ref{fig:SandPandSTotalCurrent} for $s$-wave--topological junctions and in Fig.~\ref{fig:nontopSandPandSCurrentVaryingT}(b) for $s$-wave--nontopological junctions. We remark that, while Fig.~\ref{fig:SandPandSTotalCurrent} displays currents for only a few configurations of the gap parameters, other configurations have been calculated with no important qualitative features that are distinct from those shown.

There are some key features in the Josephson currents that can reveal whether or not a junction is topological. The most prominent of these arise in the case with both energy and fermion parity relaxation: topological junctions have magnitude-preserving current switches at phase differences $\phi=n\pi$ (with $n$ integer), where the current at these switch locations is solely due to the low-energy subgap levels. Furthermore, provided that the transmission probability $\tau$ of the junction is not too close to unity, analogous magnitude-preserving switches are absent in the nontopological case [see Fig.~\ref{fig:nontopSandPandSCurrentVaryingT}(b)]. However, as can be demonstrated using Eq.~\eqref{eq:nontoptopsecular1}, nontopological variants are only able to support switchlike features at $\phi=(2n+1)\pi$, so the observation of current switches in phase increments of $\pi$ indicates the topological nature of a junction.

In addition to these qualitative signatures, our results allow a quantitative fit to experimental data using the normal-state conductance and the induced gaps as inputs, which may facilitate progress in identifying topological Josephson junctions.

\section{Numerical simulation}
\label{sec:numerics}
We now turn to testing our predictions on a lattice model\cite{ZhangKaneMele2013} for topological superconductivity in hybrid structures based on a Rashba spin-orbit coupled nanowire in proximity to an $s_\pm$-wave superconductor. The Hamiltonian consists of a left and a right subsystem, each with $N$ sites,
\begin{equation}
	H=H_{0\mathrm{L}}+H_{S\mathrm{L}}+H_{0\mathrm{R}}+H_{S\mathrm{R}}+H_{t},
\end{equation}
where $H_{0\mathrm{L}}+H_{S\mathrm{L}}$ describes the left proximitised nanowire, $H_{0\mathrm{R}}+H_{S\mathrm{R}}$ describes the right proximitised nanowire, and
\begin{subequations}
	\begin{align}
	H_{0\mathrm{L}}&=\sum_{j=1}^{N-1} \left[\left(c_{j+1}^{\dagger}V c_{j}+ \mathrm{H.c.}\right)-\mu c^\dagger_j c_j\right], \\
	H_{0\mathrm{R}}&=\sum_{j=N+1}^{2N-1} \left[\left(c_{j+1}^{\dagger}V c_{j}+ \mathrm{H.c.}\right)-\mu c^\dagger_j c_j\right],
	\end{align}
\end{subequations}
where $\mu$ is the chemical potential, $c_j=\left(\begin{array}{cc}
c_{j\uparrow} & c_{j\downarrow}
\end{array}\right)^\mathsf{T}$, and $V=-\lambda \openone_{2}-i\lambda_{\text{SO}}\sigma_{3}$ with real hopping $\lambda$ and spin-orbit energy $\lambda_\text{SO}$. The pairing is
\begin{subequations}
	\begin{align}
		\begin{split}
			H_{S\mathrm{L}}&=\frac{1}{2}e^{-i\phi/2}\left(\sum_{j=1}^{N}\Delta_{0\mathrm{L}}c_{j}^{\dagger}(i\sigma_{2})c_{j}^{\dagger}+ \mathrm{H.c.}\right.\\
			&\qquad\qquad\qquad +\sum_{j=1}^{N-1}\left.\Delta_{1\mathrm{L}}c_{j+1}^{\dagger}(i\sigma_{2})c_{j}^{\dagger}+ \mathrm{H.c.}\right),
		\end{split}\\
		\begin{split}
			H_{S\mathrm{R}}&=\frac{1}{2}e^{i\phi/2}\left(\sum_{j=N+1}^{2N}\Delta_{0\mathrm{R}}c_{j}^{\dagger}(i\sigma_{2})c_{j}^{\dagger}+ \mathrm{H.c.}\right.\\
			&\qquad\qquad\quad\left. +\sum_{j=N+1}^{2N-1}\Delta_{1\mathrm{R}}c_{j+1}^{\dagger}(i\sigma_{2})c_{j}^{\dagger}+ \mathrm{H.c.}\right)
		\end{split}
	\end{align}
\end{subequations}
and in terms of $\Delta_\pm=\Delta_{s}\pm \Delta_{p}$, this leads to
\begin{subequations}
	\begin{align}
		\begin{split}
			\Delta_{s}=\Delta_{0}-\Delta_{1}\frac{\mu\lambda}{\lambda^{2}+\lambda_{\text{SO}}^{2}},
		\end{split}\\
		\begin{split}
			\Delta_{p}=2\Delta_{1}\lambda_{\text{SO}}\frac{\sqrt{\lambda^{2}+\lambda_{\text{SO}}^{2}-\mu^{2}/4}}{\lambda^{2}+\lambda_{\text{SO}}^{2}}.
		\end{split}
	\end{align}
	\label{eq:LatticeEDPairings}
\end{subequations}
The coupling between the two subsystems is via
\begin{equation}
	H_{t}=c_{N+1}^{\dagger}\widetilde{V} c_{N} + \mathrm{H.c.}
\end{equation}
and we allow for arbitrary coupling consistent with time-reversal symmetry, i.e. we take $\widetilde{V}=-\tilde{\lambda}M$ with arbitrary $M\in \text{SU}(2)$ and $\tilde{\lambda}$ real. 

The normal-state transmission matrix $t$, via wavefunction matching, can be shown to be 
\begin{equation}
	t=\tau e^{i\chi}M,\qquad \tau=\frac{x^{2}(4-y^{2})}{(x^{2}+1)^{2}-x^{2}y^{2}},
	\label{eq:numericsTt}
\end{equation}
where $x=\tilde{\lambda}/\sqrt{\lambda^{2}+\lambda_{\text{SO}}^{2}}$ and $y=\mu/\sqrt{\lambda^{2}+\lambda_{\text{SO}}^{2}}$; the phase $\chi$, explicitly expressible in terms of $x$ and $v$,  plays no role in what follows.

This lattice model is used to numerically investigate the topological--topological and $s$-wave--topological Josephson junctions and their nontopological counterparts. In general there is excellent agreement between the scattering matrix results and the numerical lattice model. This is despite considering finite size systems and without imposing the Andreev approximation. Parameter values that were used (in units of $\lambda$) are $N=4000$, $\lambda_\text{SO}=0.15$, and $\mu=-0.7$. The corresponding value for $\tilde{\lambda}$ to produce a desired value of $\tau$ may be derived from Eq.~\eqref{eq:numericsTt}. The values for gap parameters depend on the specific junction setup and may be found in the relevant figure captions.

The numerical lattice data are plotted as black crosses and black dashed lines along with various subgap energies and continuous current contributions. For numerical simulations of topological junctions, the subgap energies display a constant line at zero energy which is due to the Majorana fermions at the wire end points away from the junction. By taking a sufficiently long nanowire, the splitting of these modes  is exponentially suppressed to zero. For the nontopological variant, there is no line at zero energy, reflecting the absence of Majorana fermions in the system.

Andreev levels for topological--topological junctions are depicted in Fig.~\ref{fig:P+SAndreevEnergies}. The numerics fit closely to the scattering matrix results, correctly reproducing the effect of $\omega$ to move the zero crossings in $\phi$-space and  the feature of subgap states escaping into the continuum. Similar agreement between the lattice model and the scattering matrix results is also achieved for the nontopological variants in Fig.~\ref{fig:NonTopP+SAndreevEnergies}. The lattice model provides the energy levels of the occupied continuous spectrum and hence the continuous contributions to the current via $I_\mathrm{cont}=\left(\pi/\Phi_0\right)\left(\mathrm{d}E_\text{cont}/\mathrm{d}\phi\right)$. These contributions are seen for topological--topological junctions in Fig.~\ref{fig:intergapcurrents1}(a)-(d) and for nontopological variants in Fig.~\ref{fig:intergapcurrents1}(e)-(h), displaying excellent agreement between the lattice model and the scattering matrix calculation.

Similarly for $s$-wave--topological junctions and the corresponding nontopological variants, the lattice model and the scattering matrix calculation are in excellent agreement. The case in which an $s$-wave admixture is included is depicted in Fig.~\ref{fig:SandP+SAndreevEnergies} for the topological case and in Fig.~\ref{fig:nontopSandP+SAndreevEnergies} for the nontopological case. The continuous contributions to the supercurrent are shown in Fig.~\ref{fig:SandPandSContContribs}(a)--(d) for the topological case and in Fig.~\ref{fig:SandPandSContContribs}(e)--(h) for the nontopological case.

\section{Conclusion}
\label{sec:concl}

We have studied the Josephson effect for time-reversal-invariant topological--topological and $s$-wave--topological junctions in the short-junction limit. In terms of the subgap energy spectrum, we found topological generalisations of Beenakker's formula [Eq.~\eqref{eq:BeenakkerFormula}] for both of the junction setups considered. For topological--topological junctions, one such expression [Eq.~\eqref{eq:energies1}] holds in the case when the topological superconductors are characterised by $\left|\Delta_s\right| \ll \left|\Delta_p\right|$. For $s$-wave--topological junctions in the analogous $\Delta_s \ll \Delta_p$ regime, three expressions are found to hold for the cases $\Delta_{0}=\Delta_{p}$ in Eq.~\eqref{eq:lowandhighenergies1}, $\Delta_{0} \gg \Delta_{p}$ in Eq.~\eqref{eq:strongconventional}, and $\Delta_0 \ll \Delta_p$ in Eq.~\eqref{eq:strongtopological}, where $\Delta_0$ is the gap parameter for the conventional $s$-wave superconductor and $\Delta_s$ and $\Delta_p$ are the respective $s$-wave and $p$-wave pairings for the topological superconductor. For more general cases we are able to find simple analytical expressions to establish key features, such as the location of zero crossings and when subgap states are lost to the continuum.

The supercurrent generally has contributions from both the subgap Andreev levels and the continuous spectrum. The continuous contributions originate from states with an energy lying in an intergap regime (arising when there is either an $s$-wave admixture present in the topological superconductor or a gap asymmetry across the junction). The continuous contributions are generally significant in magnitude relative to the subgap contributions provided that the junction is away from the tunnelling limit and that the intergap regime is not too small.

For topological--topological junctions, we found that the fractional Josephson effect can occur, which has the hallmark of a $4\pi$ periodic current. The establishment of such a current, however, is challenging due to energy relaxation. As with class D superconductors, the periodicity can be measured via the ac Josephson effect: the phase sweep may be controlled by the voltage ($\dot{\phi}=2eV/\hbar$) across the junction and the resulting periodicity of the supercurrent can be measured. The $4\pi$ periodicity of the current will only remain intact provided that the phase sweep speed is much faster than $t_r$ ($eV \gg \hbar/t_r$), the typical timescale for energy relaxation (but still slow enough so that Landau-Zener tunnelling between branches or to the continuum is avoided). For the case with energy relaxation ($eV \ll \hbar/t_r$), the current is $2\pi$ periodic. This is in contrast to class D superconductors, where the conservation of local fermion parity at the junction suffices to ensure that the $4\pi$ periodicity is robust against energy relaxation.

Due to the difficulties posed by energy relaxation in realising a $4\pi$ periodic current, we have proposed signatures unique to topological junctions in the presence of energy relaxation. For topological--topological junctions, the regime of fermion parity nonconservation is particularly useful. This is the dc Josephson effect where the zero-energy crossings of the Andreev levels [Eq.~\eqref{eq:zerocrossings}], directly linked to the topological symmetry protected energy-phase relation of Ref.~\onlinecite{ZhangKanePRB2014}, lead to a pair of current switches  in the interval $-\pi<\phi\leq \pi$. Via Eq.~\eqref{eq:zerocrossings}, the distance between the switches also provides a direct measure of the spin-orbit parameter $\omega$. For the special values  $\omega=n\pi$, the pair of  switches degenerate and can be at the same location as the switchlike features of  highly transmitting nontopological junctions. To be able to attribute the current switches to that of a topological origin, it is therefore useful either to move the junction away from such special points by changing the spin-orbit coupling or to work away from unit transparency. 

For $s$-wave--topological superconductor junctions, we also found the dc Josephson effect to be the regime with the clearest signatures of topological superconductivity. In this case, magnitude-preserving current switches occur at the time-reversal-invariant phases $\phi=n\pi$ (with $n$ integer), reflecting the existence and anomalous time-reversal properties of Majorana--Kramers pairs at these phases. In contrast, nontopological analogues may only support switchlike features for $\tau$ close to unity at ${\phi=(2n+1)\pi}$. Therefore, the observation of magnitude-preserving switches after every phase increment of $\pi$ is an indicator of topology.

We have worked in the zero temperature limit for both types of junction setup. In terms of the dc Josephson effect, our work may be straightforwardly extended to finite temperature $T$.\cite{BeenakkerPRL1991} The current switches remain intact away from zero temperature provided that ${k_\mathrm{B} T / \Phi_0 \ll I_\mathrm{s}}$ is satisfied, where $I_\mathrm{s}$ is the magnitude of the current switch. As the temperature increases outside of this regime, the switchlike features are lost as they are smoothed out over $\phi$-space.

Our work on these junctions provides a direct relationship between the excitation spectrum, the Josephson current, the normal-state conductance $G=\left(2e^2/h\right)\tau$, the various superconducting gaps, and, in the case of topological--topological junctions, the spin-orbit parameter $\omega$. Relations of this type in the nontopological case have been known to be of exceptional utility in quantitative experimental analyses of Josephson junctions.\cite{KoopsPRL1996, GoffmanPRL2000, RoccaPRL2007, BretheauNature2013, BretheauPRX2013} Our results enable the application of similar strategies to the topological case.

\begin{acknowledgements}
	This research was supported by the Royal Society, an EPSRC PhD studentship, and EPSRC grant EP/M02444X/1.
\end{acknowledgements}

\bibliographystyle{apsrev4-1}

\end{document}